\documentclass[11pt]{article}
\usepackage{epsfig,graphicx,psfrag,axodraw,bm,amssymb}
\usepackage{mathrsfs,amsfonts,hepunits}
\usepackage[hyphens]{url}\usepackage{hyperref}
\hypersetup{breaklinks=true, pagecolor=white, colorlinks=false}
\renewcommand{\theequation}{\thesection.\arabic{equation}}
\newcommand{\bear}{\begin{eqnarray}}
\newcommand{\eear}{\end{eqnarray}} 
\def\lsim{\mathrel{\rlap{\lower4.4pt\hbox{\hskip0.2pt$\sim$}}
    \raise1pt\hbox{$<$}}}
\def\gsim{\mathrel{\rlap{\lower4.5pt\hbox{\hskip0.8pt$\sim$}}
    \raise1pt\hbox{$>$}}}
\newcommand{\etaQCD}{\eta_{\mbox{\tiny \rm QCD}}}
\newcommand{\be}{\begin{eqnarray}}
\newcommand{\ee}{\end{eqnarray}}

\begin{document}

\twocolumn[ 

{\scriptsize FERMILAB-PUB-11-141-T} \\  

\vspace{-1.1cm}

\title{\bf \Large Weak-triplet, color-octet scalars and the CDF dijet excess}
\author{\large Bogdan A. Dobrescu$^1$ and Gordan Z. Krnjaic$^{1,2}$ \\ [4mm]  
\it\small  $1)$ Theoretical Physics Department, Fermilab, Batavia, IL 60510, USA\\ [2mm]  
\it\small $2)$ Department of Physics and Astronomy, Johns Hopkins University, 
Baltimore, MD 21218, USA
}


\date{\normalsize April 14, 2011; \ revised June 24, 2011}
\maketitle
 

\vspace{-.8cm}

\begin{quote}
We extend the standard model to include a weak-triplet and color-octet scalar.
This `octo-triplet' field consists of three particles, two charged and one neutral, whose
masses and renormalizable interactions depend only on two new parameters. 
The charged octo-triplet decay into a $W$ boson and a gluon is suppressed 
by a loop factor and an accidental cancellation.
Thus, the main decays of the charged octo-triplet may occur through 
higher-dimensional operators, mediated by a heavy vectorlike fermion, into quark pairs.
For an octo-triplet mass below the $t\bar{b}$ threshold, the decay into $Wb\bar{b}$ or $Wb\bar{s}$
through an off-shell top quark has a width comparable to that into $c\bar{s}$ or $c\bar{b}$.
Pair production with one octo-triplet decaying into two jets and the other decaying into a $W$ 
and two soft $b$ jets may explain the dijet-plus-$W$ excess reported by the CDF Collaboration.
Using a few kinematic distributions, we compare two mechanisms of octo-triplet pair production:
through an $s$-channel coloron and through the coupling to gluons.
The higher-dimensional operators that allow  dijet decays also lead to CP violation 
in $B_{s}\! -\! \overline B_{s}$ mixing.
\end{quote}

\vspace{.4cm}
]


\section{Introduction}

Scalar fields transforming as octets under $SU(3)_c$, the color group of the 
strong interactions, have been studied in various contexts.
The simplest type of color-octet scalar is a 
singlet under the $SU(2)_W$ group of the weak interactions, and may be referred to as an `octo-singlet'. 
These  lead to pairs of dijet resonances at hadron colliders 
\cite{Chivukula:1991zk,Dobrescu:2007yp}, and 
may explain \cite{Bai:2010dj} some deviations from the standard model predictions 
in the $3b$ search performed by the CDF collaboration \cite{HiggsPlusb}.
They also enhance the standard model Higgs boson production through gluon fusion \cite{Boughezal:2010ry}.
Octo-singlets appear as composite particles due to technicolor \cite{Hill:2002ap} and other strong-coupling
dynamics \cite{Kilic:2008pm,Bai:2010mn}, or as elementary particles in 6-dimensional extensions 
of the standard model \cite{Burdman:2006gy}
and in theories with an extended color group \cite{Bai:2010dj,Martynov:2009en}.

Weak-doublet color-octet scalars ({\it i.e.}, `octo-dou\-blets') differ dramatically from octo-singlets because the standard model gauge 
symmetry allows renormalizable couplings of octo-doublets to the standard 
quarks \cite{Manohar:2006ga}.
Only if these couplings are highly suppressed or aligned with the standard model Yukawa couplings  
can the octo-doublets be light enough to be produced at the LHC \cite{Arnold:2009ay}.
An octo-doublet field includes four color-octet states: a charged particle, a neutral one and their 
antiparticles. The hadron collider signatures of octo-doublets have been explored in
\cite{Gresham:2007ri,Gerbush:2007fe}.

In this paper we study `octo-triplets': 
real scalar fields that transform in the adjoint representation, $(8,3,0)$, of $SU(3)_c\times SU(2)_W \times U(1)_Y$.
An octo-triplet field includes three color-octet states: a particle of charge +1, its antiparticle, and 
a neutral real particle.
Akin to octo-singlets, octo-triplets are pair produced at hadron colliders through their 
couplings to gluons, and cannot decay into standard model fermions at renormalizable level
because the Yukawa couplings are not $SU(2)_W\times U(1)_Y$ invariant. 
Unlike octo-singlets, octo-triplets cannot decay into gluons unless there are additional 
fields that generate certain dimension-7 operators. 

One-loop decays of octo-triplets into a gluon and an electroweak boson are allowed,
leading to interesting collider signatures involving two gluons and two electroweak bosons. We 
will show, however, that the rate of these decays is accidentally suppressed by two orders of 
magnitude compared to usual 1-loop estimates. Thus, new heavy particles could induce the dominant
octo-triplet decay modes.  

In the presence of some vectorlike quark of mass in the TeV range, 
the charged octo-triplet may decay into a pair of standard model quarks, or into a $W$ boson and a pair of quarks 
if it is lighter than the top quark. This leads to a variety of collider signatures, including a dijet resonance, 
a $W$ boson and two softer jets. 
If the octo-triplet mass is in the 150 -- 170 GeV range, 
this signature may explain the $4.1\sigma$  excess observed by the CDF Collaboration in the dijet resonance plus $W$ 
final state \cite{CDF7, Aaltonen:2011mk}\footnote{The D0 search in the same channel \cite{Abazov:2011af} has a larger background 
and less data, so that it might not be sensitive enough to the signature proposed here.}. 
Some alternative explanations can be found in \cite{Bai:2010dj},\cite{Yu:2011cw}-\cite{Sato:2011ui}.  
At the LHC, octo-triplets with much larger masses ($\sim$1 TeV) may be probed in several final states.
    
Octo-triplets may be elementary particles ({\it e.g.}, part of the 75 representation of $SU(5)$ grand unification), 
or may arise as composite ones, for example 
as fermion-antifermion bound states \cite{Bai:2010mn}. We treat the octo-triplets as point-like particles, 
which is a good approximation only when the compositeness scale is substantially higher than the 
octo-triplet mass.

In Section 2 we analyze the extension of the standard model by one real octo-triplet field. Section 3.1 
introduces a heavy vectorlike quark which mediates octo-triplet decays into standard model quarks. 
Section 3.2 discusses flavor-changing processes. The Tevatron phenomenology of charged octo-triplets is 
explored in sections 3.3 (QCD pair production) and 3.4 (resonant pair production). The predictions for LHC
are discussed in section 3.5. Our conclusions are summarized in Section 4. In the appendices we present the Feynman rules for octo-triplets, 
and then we compute the rates for the 3-body weak decay of the charged octo-triplet and for the 2-body 1-loop 
decays of color-octet scalars.
  
\section{Octo-triplet scalar} 
\label{sec:triplet}
\setcounter{equation}{0}

We consider the standard model plus an octo-triplet, $\Theta^{a\alpha}$, which is a real field of spin 0 transforming as $(8,3,0)$
under the standard $SU(3)_{c} \times SU(2)_W \times U(1)_Y$ gauge group.
We use indices from the beginning of the Roman and Greek alphabets
to label the $SU(3)_{c}$ and $SU(2)_W$ generators, respectively: $a,b,c = 1, \ldots ,8$ and  $\alpha,\beta,\gamma = 0,1,2$.

\subsection{Interactions and masses}

All interactions of the octo-triplet with standard model gauge bosons are contained in the 
kinetic term 
\be
\frac{1}{2} \left(D_{\mu} \Theta^{a\alpha} \right)\left(D^{\mu}\, \Theta^{a\alpha} \right)  ~~,
\label{kinetic}
\ee
where $\mu$ is a Lorentz index and the covariant derivative is given by
\be 
\hspace*{-2mm}
D_\mu \Theta^{a\alpha} \! = \! \partial_\mu \Theta^{a\alpha} \! + \! g_s f^{abc} G_\mu^b \Theta^{c\alpha}  
\! + \! g \epsilon^{\alpha\beta\gamma} W_\mu^\beta \Theta^{a \gamma}  .
\ee
Here $f^{abc}$ and $\epsilon^{\alpha\beta\gamma}$ are the totally antisymmetric tensors of the $SU(3)_{c}$ and $SU(2)_{W}$ groups, respectively,
$g_{s}$ and $g$ are the $SU(3)_{c} \times SU(2)_{W}$ gauge couplings, $G_\mu^a$ is the gluon field, and $W_\mu^\alpha$ is the weak gauge field.
The octo-triplet field includes three particles: an electrically-neutral color-octet real scalar $\Theta^{a 0}$, a color-octet scalar
of electric charge +1, $\Theta^{a +}$, and its antiparticle $\Theta^{a -}$:
\be
\Theta^{a \pm} = \frac{1}{\sqrt{2}} \left(\Theta^{a 1} \mp i \Theta^{a 2} \right) ~~.
\ee
When referring informally to the octo-triplet particles we use the  $\Theta^{\pm}$ and  $\Theta^{0}$ symbols without 
displaying the color index $a$.

The kinetic term (\ref{kinetic}) includes interactions of the $W$ boson with two octotriplet particles,
\be
- i g W_\mu^- \left[ (\partial_\mu \Theta^{a + }) \Theta^{a 0}  - \Theta^{a + } \partial_\mu \Theta^{a 0 } \right] + {\rm H.c.}  ~~,
\label{triplet-bosons}
\ee
and also with an additional gluon:
\be
 2 i g g_s f^{abc} \,  G^{\mu a} \left(W_\mu^+ \Theta^{b -} - W_\mu^- \Theta^{b +} \right) \Theta^{c 0} ~~.
 \label{triplet-gluons}
\ee
Similar interactions involve a $Z$ boson and two octo-triplet particles of the same charge, with or without an additional gluon.
The interactions of like-sign octo-triplets with one or two gluons (photons) are completely specified by QCD (QED) gauge invariance.
The Feynman rules for octo-triplets are given in Appendix A. 

The mass of the octo-triplet arises from two terms in the Lagrangian:
\be
- \frac{1}{2} \left( M_0^2 - \lambda_{H} H^\dagger H\right) \Theta^{a \alpha} \Theta^{a \alpha} ~~,
\label{massterms} 
\ee
where $\lambda_{H}$ is a real dimensionless parameter.
The VEV of  the standard model Higgs doublet $H$ 
has a value $v_H \simeq 174$ GeV,  so that the mass of the octo-triplet field is 
\be
M_\Theta = \sqrt{M_0^2 - \lambda_{H}v_H^2} ~~.
\ee
We require $M_\Theta > 0$ ({\it i.e.,} $\Theta^{a \alpha}$ does not acquire a VEV) 
in order to preserve $SU(3)_{c}$ gauge invariance. 

The commutation relations of the Pauli matrices $\sigma^{\alpha}$  imply that other 
operators contributing to the octo-triplet mass, such as 
$(H^{\dagger} \sigma^{\alpha}\sigma^{\beta} H) \Theta^{a \alpha} \Theta^{a \beta}$, are either 
identical to the last one in Eq.~(\ref{massterms}) or vanish.
Thus, at tree level $\Theta^{\pm}$ and $\Theta^{0}$ are degenerate states, having 
masses equal to $M_\Theta$.
At one loop, the electroweak interactions break this degeneracy. 
The mass splitting between the charged and neutral octo-triplets 
is \cite{Dodelson:1991iv}
\be
\delta M \equiv M_{\Theta^+} -  M_{\Theta^0} \simeq 
\frac{1 - \cos\theta_W }{2 \sin^2\!\theta_W} \alpha M_W  
\label{mass-splitting}
\ee 
up to corrections of order $(M_W/M_\Theta)^2$.
We will see shortly that the octo-triplets have lifetimes much longer than the QCD scale, so that 
they hadronize. The lightest physical states are ``octo-hadrons'' given by a
$\Theta^{0}$ or $\Theta^\pm$ bound to gluons or quark-antiquark pairs. The mass difference 
$\delta M$ between the lightest charged and neutral octo-hadrons is of the same sign and 
order of magnitude as $M_{\Theta^+} -  M_{\Theta^0}$, so that $\delta M \sim 0.2$ GeV.

$SU(2)_W \times U(1)_Y$ gauge-invariance forbids any renormalizable interaction of 
the octo-triplet with standard model fermions.
The most general renormalizable Lagrangian (${\cal L}_\Theta$) for the octo-triplet scalars is given by
the kinetic term (\ref{kinetic}), the potential terms quadratic in $\Theta$ given in Eq.~(\ref{massterms}), 
as well as a cubic term and quartic terms:
\be  \mu_\Theta f^{abc} \epsilon^{\alpha\beta\gamma}  \Theta^{a\alpha} \Theta^{b\beta} \Theta^{c\gamma}
-  \lambda_{\Theta} \left(\Theta^{a\alpha}\Theta^{a\alpha} \right)^2    ~~,
\label{eq:lag}
\ee
where $\lambda_\Theta > 0$ is a dimensionless parameter, and for simplicity we display only one quartic term.
The mass parameter  $\mu_\Theta$ may be positive or 
negative, but its size should not be larger than $O(M_\Theta \lambda_\Theta^{-1/2})$ in order to prevent a $\Theta$ VEV.
The above cubic term gives the following interaction among the charged and neutral octo-triplet particles:
\be
2 i \mu_\Theta f^{abc} \Theta^{a +} \Theta^{b -} \Theta^{c 0}   ~~.
\label{triplet-cubic}
\ee


\subsection{ Collider signals of octo-triplets}  

In the $\mu_\Theta \to 0$ limit, the Lagrangian ${\cal L}_\Theta$ has an accidental $\mathcal Z_{2}$ symmetry 
that makes the lightest octo-triplet ({\it i.e.,} $\Theta^{0}$) stable. 
The charged octo-triplet decays at tree level into $\Theta^{0}$ and an off-shell $W$ boson.
Computing the 3-body width to leading order in $\delta M$ (Appendix B) we find
\be
\Gamma \!\left(\Theta^\pm \!\to\! \Theta^{0} e^\pm \nu \right) \simeq \frac{\alpha^2}{15\pi\sin^4\!\theta_W \!}
\frac{ \! (\delta M)^{5}\!\! }{M_W^4}  ~. 
\label{3width}
\ee 
Given the small mass splitting $\delta M \sim 0.2$ GeV [see the comment after Eq.~(\ref{mass-splitting})],
the only other relevant decay mode is
$\Theta^\pm \! \to\! \Theta^{0} \mu^\pm \nu$ with a decay width further phase-space suppressed compared to 
Eq.~(\ref{3width}). Hence, the total tree-level width of $\Theta^\pm$ is $\Gamma_{\rm tree}(\Theta^\pm) \simeq 1.8\times \!10^{-16}$ GeV.
This 3-body decay width corresponds to a decay length of $1.1 \,{\rm cm}.$

For $\mu_\Theta \neq 0$, the charged octo-triplet decays into gauge bosons at one loop, with  $Wg$ being 
the only 2-body final state allowed by charge conservation. The diagrams responsible for this decay are shown in 
Figure \ref{fig:triangle}. The computation of the decay width described in Appendix C 
gives
\be
\hspace*{-4mm}  \Gamma\! \left(\Theta^\pm\! \to \!W^\pm g\right) \simeq \!\frac{ \alpha \alpha_s \mu_\Theta^2}{\pi^3 \sin^2\!\theta_W M_\Theta} 
f\!\left(M_W/M_\Theta\right) ,
\label{eq:triangle}
\ee
where the function $f(R)$ is defined in Eq.~(\ref{function}). For $M_\Theta$ varying between 150 GeV and 1 TeV, $f(R)$ grows from $(4.0 -10.3) \times 10^{-3}$,
corresponding to  
$\Gamma\! \left(\Theta^\pm\!\! \to \!W^\pm g\right)$ in the $(4.3 - 11.2) \times 10^{-7}\! \mu_\Theta^2/M_\Theta$ range. While one
might naively expect $f(R)$ to be of order one, this function is accidentally suppressed: $f(0)\!\propto\! 
 \left(\pi^{2}/9 - 1\right)^{2} $ as shown in Eq.~(\ref{flimit}), while for larger values of $R$, the function decreases further due to phase space suppression.

\begin{figure}[b!]
\vspace*{-0.1cm}
\begin{center} 
{
\unitlength=0.5 pt
\SetScale{0.5}
\SetWidth{2.}      
\normalsize    
{} \allowbreak
\thicklines
\begin{picture}(100,100)(20,-100)
\DashLine(15,30)(70,30){4}
\DashLine(70,30)(140,70){4}
\DashLine(70,30)(140,0){4}
\DashLine(140,0)(140,70){4}
\Gluon(140,0)(199,0){-4}{5}
\Photon(140,70)(199,70){4}{6}
\Text(17,13)[c]{\small $\Theta^+$}
\Text(100,67)[c]{\small $\Theta^+$}\Text(100,0)[c]{\small $\Theta^0$}
\Text(158,30)[c]{\small $\Theta^0$}
\Text(208,0)[c]{\small  \ $g$}\Text(221,70)[c]{\small  $W^+$}
\end{picture}
\quad\quad
%
\begin{picture}(100,100)(-78,-100)
\DashLine(15,30)(70,30){4}
\DashLine(70,30)(140,70){4}
\DashLine(70,30)(140,0){4}
\DashLine(140,0)(140,70){4}
\Gluon(140,0)(199,0){-4}{5}
\Photon(140,70)(199,70){4}{6}
\Text(17,13)[c]{\small $\Theta^+$}
\Text(100,67)[c]{\small $\Theta^0$}\Text(100,0)[c]{\small $\Theta^+$}
\Text(158,30)[c]{\small $\Theta^+$}
\Text(208,0)[c]{\small  \ $g$}\Text(221,70)[c]{\small  $W^+$}
\end{picture}
\quad\quad\quad
%
\begin{picture}(100,200)(200,10)
\DashLine(15,30)(70,30){4}
\DashCArc(105,30)(35,0,180){4}
\DashCArc(105,30)(35,180,0){4}
\Photon(140,30)(195,60){4}{6}
\Gluon(140,30)(195,0){4}{5}
\Text(17,13)[c]{\small $\Theta^+$}
\Text(70,67)[c]{\small $\Theta^+$}\Text(70,-6)[c]{\small $\Theta^0$}
\Text(207,0)[c]{\small  \ $g$}\Text(220,60)[c]{\small  $W^+$}
\end{picture}
}
\end{center}
\vspace*{0.1cm}
\caption{Charged octo-triplet decay to a $W$ boson and a gluon.
}
\label{fig:triangle}
\end{figure}
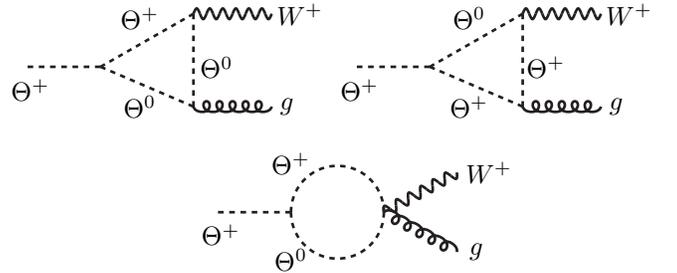

The neutral octo-triplet also decays at one loop, into a gluon and $Z$ boson or photon, with partial widths 
\bear
&& \hspace*{-14mm} \Gamma\! \left(\Theta^0\! \to \! Z g\right) \simeq \!\frac{ \alpha \alpha_s \mu_\Theta^2}{\pi^3 \tan^2\!\theta_W M_\Theta} 
f\!\left(M_Z/M_\Theta\right)~~,
\nonumber \\
&& \hspace*{-14mm} 
 \Gamma\!\left(\Theta^{0} \to \gamma g \right)  \simeq \frac{ \alpha \alpha_s \mu_\Theta^2}{\pi^3 M_\Theta} 
f\!\left(0\right) ~ .
\label{eq:triangleZ}
\eear
When $M_\Theta$ varies between 150 GeV and 1 TeV, the branching fraction for $\Theta^{0} \to \gamma g$ 
decreases from 53\% to 24\%.
The decay $\Theta^{0} \to gg$ does not occur at one loop due to $SU(2)_W$ invariance (this decay requires 
a dimension-7 operator involving two Higgs fields).

\begin{figure}
\begin{center}%
\hspace*{-3.4cm}
\parbox{2.2in}{\includegraphics[width=.48 \textwidth]{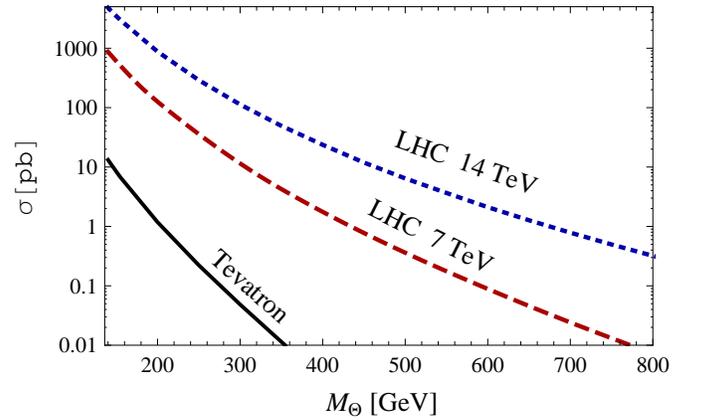}}\caption{Leading order cross section for charged octo-triplet pair production at the Tevatron (black solid line) and 
LHC at $\sqrt{s} = 7\,\TeV$ (red dashed line)  and  at $\sqrt{s} = 14 \,\TeV$ (blue dotted line).}
\label{fig:cross-section}
\end{center}
\end{figure}

At hadron colliders, octo-triplets are copiously pair produced due to their QCD couplings to gluons.
The rate for $\Theta^0\Theta^0$ production is the same as for an octo-singlet of same mass \cite{Dobrescu:2007yp,Bai:2010dj},
while $\Theta^+\Theta^-$ production is twice as large (additional contributions due to photon and $Z$ exchange are negligible). 
In Figure \ref{fig:cross-section}  we show the leading order $\Theta^{+}\Theta^{-}$ 
production cross section at the Tevatron and LHC, computed with MadGraph 5 \cite{Alwall:2011uj}
(with model files generated by FeynRules \cite{Christensen:2008py})
using the CTEQ 6 parton distribution functions \cite{Pumplin:2002vw}.  The QCD corrections are not included
in this plot; we expect their inclusion to shift these curves upwards by $O(50\%)$.

Note that single octo-triplet production (through diagrams similar to those in Figure \ref{fig:triangle}) 
is negligible because it is suppressed by a loop factor, the weak coupling constant, and $(\mu_\Theta/M_\Theta)^2$.

The $\Theta^0\Theta^0$ pair leads to $(Z j)(Z j)$, $(\gamma j)(Z j)$ and $(\gamma j)(\gamma j)$ final states, where 
$j$ is a gluonic jet and the parantheses indicate that the two objects form a resonance of mass $M_\Theta$.
The  $\Theta^+\Theta^-$ pair leads to $(W^{+} j)(W^{-} j)$ final states, unless $\mu^{2}_\Theta/M_\Theta \lesssim O(10^{-9})$ GeV 
which leads to a
large branching fraction for the $\Theta^\pm \to \Theta^0 e^\pm \nu$ decay. This latter case gives the same final state as in
$\Theta^0\Theta^0$ production because the electron and neutrino are very soft and most likely do not pass the cuts (even when $\Theta^\pm$ is 
boosted the electron is not isolated). If  $\mu^{2}_\Theta/M_\Theta \simeq 4 \times 10^{-10}\,\GeV$, then the 2- and 3-body decays of  $\Theta^\pm$ have comparable widths, 
so that the $\Theta^+\Theta^-$ pair leads to $(W j)(Z j)$ and $(W j)(\gamma j)$ final states, with the $(Z j)$ and $(\gamma j) $ vertices 
originating from  displaced vertices.

Let us briefly discuss the $(W^{+} j)(W^{-} j)$ signal at the LHC, where the production cross section, Figure\,\ref{fig:cross-section},
can be very large. This final state is most easily identified when one $W$ decays leptonically while the other decays
hadronically with an overall $W+4j$ signature. The $W$ decay products reconstruct $W$ resonances and can thereby be isolated from
the other jets in the event.  These remaining jets can then be grouped alongside the known $W$ decay products and used to reconstruct
pairs of octo-triplet resonances. Backgrounds to this signature include $W+$ jets production and  $t\bar t$ pair-production,
the latter of which can be substantially reduced by anti-$b$ tagging.

Besides nonresonant pair production, octo-triplet scalars may induce resonant signatures at hadron colliders because, 
like other long-lived colored particles  \cite{ Kats:2009bv,Kim:2008bx}, they form bound states. 
If the octo-triplet width is much less than the binding energy $E_{\mathcal B}$ due to gluon exchange, 
then bound states form before either particle
 decays.
This is the case for all values of $M_{\Theta}$ and $\mu_{\Theta}$, as the dominant 2-body octo-triplet width given in Eq. (\ref{eq:triangle}) 
easily satisfies
\be
\Gamma(\Theta^{\pm }\to W^{\pm}g) \ll E_{\mathcal B} = \frac{9}{4}\alpha_{s}^{2} \,M_{\Theta}~,~
\ee
  For simplicity, we consider only the 
 formation of color-singlet bound states $\cal B$; 
 this is the dominant channel and our qualitative conclusions 
 apply to different color-representations of bound states.

 The bound states annihilate into gauge-boson pairs before either constituent decays. 
 Bound states of neutral octo-triplets can only annihilate through 
 the processes $ \Theta^{0} \Theta^{0} \!\to \!{\cal B}\!\to\! gg, W^{+}W^{-}$, while
 the annihilation of bound states of charged octo-triplets yields a rich variety of vector boson pairs:
  $\Theta^{+} \Theta^{-} \!\!\to {\cal B}\to gg, W^{+}W^{-}, ZZ, $ $\gamma\gamma$, $\gamma Z$. 

Given that the bound-state effects are mainly due to gluon exchange, the production of the 
$\Theta^{+} \Theta^{-}$ bound state is approximately equal to that of octo-singlets computed in 
\cite{Bai:2010mn}: for $M_{\Theta}\! \approx 150 \,\GeV$ the Tevatron cross section is
 $\sigma(p\bar p \to{ \cal B}) \simeq O(100) \,\mathrm{ fb}$.
 Since the width $\Gamma(\Theta^{\pm }\!\!\to W^{\pm}g)$ is orders of magnitude 
 smaller than the main channel for bound states
   $\Gamma({\mathcal B}\to gg)  \simeq 0.04 \,\GeV$ \cite{Bai:2010mn}, 
the annihilation dominates and yields dijet resonances with invariant mass
   $M_{\cal B} = 2M_{\Theta} - E_{\cal B}$. This cross section is too small to be observed at the Tevatron.
  
  At the LHC, the bound state production cross-section can be considerably  larger.  
For $M_{\cal B} \sim 1 \,\TeV$, the cross section is 
  $\sigma( pp \to{ \cal B}) \sim$  $O(1 \,\mathrm{pb})$
 \cite{Kim:2008bx}  at $\sqrt{s} = 14 \,\TeV$, which might allow the annihilation 
 signal to compete with the QCD background
  and give an observable resonance. The   electroweak diboson channels
   are suppressed relative to $gg$, but give cleaner signals, which contribute to
  standard Higgs searches.  
  
  While the above discussion has been limited to the dominant color-singlet
  bound state, octo-triplets can also form bound states 
  in higher color representations with exotic annihilation signatures. 
  For instance, color-octet bound states annihilate into either $\gamma g$ or $Z g$ regardless of whether 
  the bound state comprises charged or neutral scalars.


\section{Octo-triplet decays via higher-dimensional operators}
\setcounter{equation}{0}

Since the octo-triplet widths in Eqs.~(\ref{3width})-(\ref{eq:triangle}) are tiny, higher-dimensional operators induced 
at the TeV scale could lead to other decays with substantial branching fractions.

Dimension-5 operators allow the coupling of an octo-triplet to a pair of standard model quarks 
involving a derivative,
\be
\frac{c_{ij} }{m_{\psi}} \,\Theta^{a\alpha} \,\overline{Q}_L^{\,i}\,T^a \,\frac{\sigma^\alpha}{2}  \gamma^\mu D_\mu Q_L^{\,j}  + {\rm H.c.}~~,
\label{eq:leff}
\ee
or in the presence of the Higgs doublet,
\be
\Theta^{a\alpha} \,\overline{Q}_L^{\,i} T^a \frac{\sigma^\alpha}{2} 
\left( \frac{c_{ij}^\prime}{m_\psi} \widetilde{H} u_R^j+ \frac{c_{ij}^{\prime\prime}}{m_\psi}{H} d_R^j \right) ~~.
\label{eq:leffH}
\ee
$Q_L^i$, $u_R^j$ and $d_R^j$ are the quark fields in the gauge eigenstate basis; $i,j = 1,2,3$ label the fermion generation; $\sigma^{\alpha}$ is a 
Pauli matrix; $m_{\psi}$ is the mass of some heavy
field that has been integrated out; and  $c_{ij}$,  $c^\prime_{ij}$ and  $c^{\prime\prime}_{ij}$ are dimensionless coefficients. Using the field
equations,  one can replace the last operator, with coefficient $c^{\prime\prime}_{ij}$, by a linear transformation (involving the standard model 
Yukawa couplings) of the $c_{ij}$ and $c^{\prime}_{ij}$ coefficients.

\subsection{Octo-triplet plus a vectorlike quark}

The dimension-5 operators (\ref{eq:leff}) and (\ref{eq:leffH}) can be induced, for example, by a heavy vectorlike quark $\Psi$ that transforms as $( 3,2,1/6)$ under
$SU(3)_{c}\times SU(2)_{W}$ $\times U(1)_{Y}$, {\it i.e.} the same way as SM quark doublets  $Q_{L}^{i}$.
Renormalizable interactions of $\Psi$ with the octo-triplet, 
\be  \hspace*{-4mm}
{\cal L}_{\Theta\Psi} = \Theta^{a\alpha}\,\overline \Psi_R \,T^{a} \frac{\sigma^{\alpha}}{2} \left( \eta_i  Q_L^i
+ \eta_{\psi}  \Psi_L \right)  +  {\rm H.c.}   ~,
\label{eq:newint}
 \ee
 and with the Higgs doublet,
 \be
{\cal L}_{H\Psi} = 
\overline{\Psi}_L  \left( \lambda_i^u \, H u_{R}^i+ \lambda_i^d \, \widetilde{H} d_{R}^i \right) ~~,
 \ee
are allowed. Here $\eta_i$, $\eta_\psi$, $\lambda^u_i$ and $\lambda^d_i$ are dimensionless couplings
and $ \widetilde{H} = i\sigma^{2} H^{\dagger}$.
Gauge-invariant fermion mass terms are also allowed:
\be 
 - m_{\psi} \overline \Psi_L \Psi_R - \mu_{i}  \overline{Q}_L^{\,i}\Psi_{R} +  {\rm H.c.} 
 \ee
For $m_\psi \gg M_\Theta$, the $\Psi$ fermion can be integrated out, giving rise to the operators (\ref{eq:leff}) 
through the ${\cal L}_{\Theta\Psi} $ interactions,
and to the operators (\ref{eq:leffH}) through a combination of ${\cal L}_{\Theta\Psi} $ and ${\cal L}_{H\Psi}$ interactions.


\begin{figure}[t]
\begin{center}
\unitlength=0.7 pt
\SetScale{0.7}
\SetWidth{0.9}      
\normalsize    
{} \allowbreak
\begin{picture}(100,90)(130,0)
\DashArrowLine(15,30)(70,30){4}
\ArrowLine(110,50)(70,30) 
\Line(103,53)(117,48)
\Line(114,58)(107,43)
\ArrowLine(142,67)(109,50)
\ArrowLine(70,30)(135,-2) 
\Text(22,16)[c]{\small $ \Theta^+$}
\Text(78,54)[c]{\small $ \overline{\Psi^d_R} $}
\Text(160,71)[c]{\small $\overline{s}_L $}
\Text(151,-4)[c]{\small $c_{L}$}
\put(181,0){
\DashArrowLine(15,30)(70,30){4}
\ArrowLine(135,58)(70,30) 
\Line(103,6)(117,13)
\Line(112,2)(109,17)
\ArrowLine(110,10)(142,-7)
\ArrowLine(70,30)(110,10) 
\Text(22,16)[c]{\small $ \Theta^+$}
\Text(78,9)[c]{\small $ \Psi_R^u $}
\Text(152,60)[c]{\small $\overline{s}_L $}
\Text(159,-8)[c]{\small $c_{L}$}
}
\end{picture}
\end{center}
\vspace*{-0.2cm}
\caption{Charged octo-triplet decay to quarks in the presence of a vectorlike 
quark doublet $\Psi = (\Psi^u,\Psi^d)$. 
The mass mixing of $\Psi$ with the standard model quarks is depicted by $\times$. 
Similar diagrams lead to the decay of the neutral $\Theta^0$ scalar into quark pairs. }
\label{fig:decay}
\end{figure}
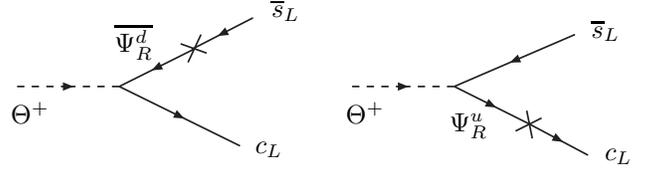

Let us assume for simplicity that $\lambda^u_i$ and $\lambda^d_i$ are negligible, and that the mass mixing parameters
satisfy $\mu_i \ll m_\psi$. In this case the 
coefficients $c_{ij}$  can be computed in the mass insertion approximation:
\be
c_{\,ij} =  - \frac{ i \, \eta_{i}^{*} \mu_{j}}{m_{\psi}}~~.
\label{eq:coef}
\ee
These are the coefficients at the scale $m_{\psi}$; running down
from $m_{\psi}$ to $M_{\Theta}$ may change $c_{\, ij}$ at $M_{\Theta}$ by
an $O(1)$ factor, which we can absorb into the definition of $\eta_{i}$.

Using the quark field equations, we find that Eq. \!(\ref{eq:leff}) contains the following 
interactions between $\Theta^+$ and the mass-eigenstate quark fields $U^i$ and $D^j$: 
\bear
&& \hspace*{-1.6cm} \frac{-i}{ \sqrt{2} m_{\psi}\!} \, \Theta^{a +} \, \overline{U}^{\,i} T^a \!
\left[  \left( C\, V_{\rm KM}\right)_{ij} m_{d_j} P_R  \right.
\nonumber \\  
&& \hspace*{.1cm}
- \left.  m_{u_i} \left(   C^\dagger \, V_{\rm KM} \right)_{ij} P_L \right] D^j   + {\rm H.c.} 
\label{int-mass-basis}
\eear
where $V_{\rm KM}$ is the CKM matrix, and $m_{u_i}$, $m_{d_i}$ are the physical masses for the 
quarks of the $i$th generation. The  $3\times 3$ matrix $C$ is given by 
\be
C = V_{u_L}^\dagger c \, V_{u_L} ~~,
\ee
where $c$ is the matrix whose  elements are given in Eq.~(\ref{eq:coef}),
and $V_{u_L}$ is the matrix that transforms the left-handed 
up-type quarks from the mass eigenstates to the gauge eigenstates, $u = V_{u_L} U$.

Based on interactions (\ref{int-mass-basis}) we find 
that the width for the decay of the charged octo-triplet into a quark pair is
\be 
\!\!\Gamma(\Theta^+ \rightarrow c \,\bar{s}) \simeq  \frac{ m_{c}^{2} + m_{s}^{2 } }
{ 64 \,\pi\, m_{\psi}^2}   \, |C_{22}|^2  M_{\Theta}~, 
\label{quarkwidth}
\ee
where we have omitted ${\mathcal O(m_{q}^{4})}$ terms and off-diagonal CKM elements, 
and have not included QCD corrections. 
Taking the charm quark mass $m_{c} = 1.3$ GeV gives 
\be \hspace*{-1cm}
\Gamma(\Theta^+ \! \rightarrow c \,\bar{s}) \!\! & \simeq&  \!\!\!  1.3\times10^{-6}  \, \GeV \;
|C_{22}|^2 
  \nonumber \\ [2mm]
&\times& \!\!\!\!    \frac{M_{\Theta}}{150 \,\GeV}\! \left( \frac{1\,\TeV}{m_{\psi} }\right)^{\!2}\!~ .  
\vspace*{-0.8 cm}
\label{quarkwidthnumbers}
\ee
Compared with the decay into $W g$ computed in Eq.~(\ref{eq:triangle}), the above $\Theta^+$ decay into a pair of jets 
can easily dominate. For example, for $M_{\Theta} = 150$ GeV, $\mu_{\Theta} = 1$ GeV, 
$C_{22} = 0.1$, and $m_{\psi} = 1.1$ TeV, 
we find $\Gamma(\Theta^+ \! \rightarrow c \,\bar{s})  \simeq 3.7 \, \Gamma(\Theta^+ \! \rightarrow W \,g)$.  

The width for the decay into $c \,\bar b$ is sensitive to different $C_{ij}$ parameters:
\be \hspace*{-0.5cm}
\frac{\Gamma (\Theta^+ \! \! \rightarrow c \,\bar{b})}{\Gamma ( \Theta^+ \! \! \rightarrow c \,\bar{s})} 
\simeq \frac{1}{|C_{22}|^2 } \left(\frac{m_b^2}{m_c^2}  \, |C_{23}|^2 + |C_{32}|^2 \!\right) ,
\label{ratio-bc}
\ee
where $m_{b} \approx 4.2$ GeV is the $b$ quark mass. 

If $M_{\Theta} > m_{t} + m_{b}$, the decay involving a top quark opens up:
\be \hspace*{.1cm}
 \Gamma(\Theta^+ \!\! \rightarrow \! t \bar{b}) \!\!\! & \simeq &  \!\!\!  2.2\times \! 10^{-2}  \, \GeV \;
|C_{33}|^2 \left(\!1\! - \!\frac{m_t^2}{M_{\Theta}^2} \right)^{\! 2}
  \nonumber \\ 
&\times& \!\!\!\!   \frac{ M_{\Theta}}{150 \,\GeV} \! \left( \frac{1\,\TeV}{m_{\psi} }\right)^{\!2}\!~ ,  
\label{tbbar}
\ee 
where we have set $m_{t} = 173$ GeV and ignored $m_b^2$ terms. 
Due to an $m_t^2/m_c^2$ enhancement compared to Eq.~(\ref{quarkwidth}), this decay dominates
unless $|C_{33}| <  10^{-2} |C_{22}|$.
The decay $\Theta^+ \!\! \rightarrow \! t \bar{s}$ has the same width except for the $C_{33} \to C_{23}$ replacement.

Similar expressions give the 2-body widths for the neutral octo-triplet decaying to $c\bar{c}$, $b\bar{b}$, or top pairs 
if $M_{\Theta} > 2 m_t$.

The  $m_t^2/m_c^2$ enhancement in Eq.~(\ref{tbbar}) is so large that even for $M_{\Theta} < m_{t} + m_{b}$ 
the 3-body decay through an off-shell top quark, 
$\Theta^+ \! \rightarrow W^+ b \,\bar{b}$, needs to be taken into account. Its width is 
\be \hspace*{-.4cm}
\Gamma(\Theta^{+}\!\!\to\!  W^+ b\bar{b}) = 
\frac{\alpha \,  |C_{33}|^2\,  m_{t}^4 \,}{64 \pi^2 \sin^2\!\theta_W \, m_\psi^2}  {\cal F}(M_\Theta)~.
\ee
The the function ${\cal F}$, of mass dimension $-1$, is given by integrating the matrix element over phase space:
\bear
 &&\hspace*{-1.8cm} {\cal F}(M_\Theta) =  \int_{0}^{E_0} 
\!\!d\overline E_{\bar{b}} \! \int_{E_0 - \overline E_{\bar{b}}}^{E_b^{\rm max}} \!\!\!\!\!
dE_b   
\nonumber \\ && \hspace*{-1.cm} 
 \frac{        \! E_b \!+ (E_0 \!-\!   \overline  E_{\bar{b}})\!\left[ \frac{2 M_{\Theta}}{M_W^2} (E_0 \!-\! E_b )\! - \!1\right]    }{ 
(M_{\Theta}^{2} - 2 M_{\Theta} \overline E_{\bar{b}} - m_{t}^{2} + m_b^2 )^{2}  + m_{t}^{2} \Gamma_{t}^{2} } ~,
\label{offshell}
\eear
where $E_0$ is the maximum  energy of the $\bar{b}$ or $b$ jet,
\be
E_0 = \frac{M_\Theta^2 \!- M_W^{2}}{2 M_\Theta} ~~, 
\ee
and $E_b^{\rm max}$ is the maximum $b$ energy for a fixed  $\bar{b}$ energy $\overline E_{\bar{b}}$,
\be
E_b^{\rm max} = \frac{E_0\! - \overline E_{\bar{b}}}{ 1 - 2\overline E_{\bar{b}}/M_{\Theta}}  ~~.
\ee
In Eq.~(\ref{offshell}) we neglected $m_b$ everywhere (which is a good approximation for 
$m_b^2 \ll E_0^2$) with the exception of the denominator where the $m_b^2$ term becomes important for $M_{\Theta}$ near the 
2-body threshold, $m_{t} + m_{b}$. To cover  that  case we also included the top quark width, $\Gamma_t \approx 1.3$ GeV, in the propagator.
Numerically, the 3-body width can be written as
\be 
\hspace*{-1.cm} 
\Gamma(\Theta^{+}\!\!\!\to\!  W^+ b\bar{b}) \!\! & \simeq&  \!\!\!  2.9 \times 10^{-6}  \, \GeV \;
|C_{33}|^{2}
  \nonumber \\ 
&& \hspace*{-0.7cm} \times \frac{ {\cal F}(M_\Theta)}{ {\cal F}(150 \; \GeV)\!} \! \left( \frac{1\,\TeV}{m_{\psi} }\right)^{\!2}\! .  
\label{tbbarvirtual}
\ee 
The ratio ${\cal F}(M_\Theta)/{\cal F}(150 \; \GeV)$ is given by 1.51 for $M_\Theta = 155$ GeV, and 
by 2.28 for $M_\Theta = 160$ GeV.

It is remarkable that the above 3-body decay through a virtual top quark has a width close to that for the 2-body decay into $c\bar s$,
given in Eq.~(\ref{quarkwidthnumbers}).
Assuming for illustration that $|C_{23}|, |C_{32}| \ll |C_{22}| = |C_{33}|$
we find that the branching fraction into  $W b\bar{b}$ is  69, 76, 82\% for $M_\Theta = 150,155, 160$ GeV,
respectively.

Finally, the decay $\Theta^+ \!\! \rightarrow \! W b  \bar{s}$, of width 
\be \hspace*{-0.7cm} 
\Gamma(\Theta^{+}\!\!\!\to\!  W^+ b\bar{s})  \approx \frac{|C_{23}|^{2}}{|C_{33}|^{2}} \, \Gamma(\Theta^{+}\!\!\!\to\!  W^+ b\bar{b})  ~,
\ee
may also have a substantial branching fraction if the $C_{23}$ parameter is large. In that case, though, the main competing channel 
is likely to be $\Theta^+ \! \! \rightarrow c \,\bar{b}$, as can be seen from Eq.~(\ref{ratio-bc}).

\subsection{$B_s - \bar{B}_s$ mixing}

\begin{figure}[t]
\begin{center}
{
\unitlength=0.7 pt
\SetScale{1.0}
\SetWidth{0.9}      
\normalsize    
{} \allowbreak
\begin{picture}(100,100)(10,56)
\put(-0,130){

\Text(0,14)[c]{\small $\overline b_L$}
\ArrowLine(  25  ,0)(-10,0)
\ArrowLine(50,0)(  25  ,0)

\Text(53, -33)[c]{\small $\Theta^0$}
\DashLine(50, -45)(50, 0){4}

\Text(5,-78)[c]{\small $s_L$}
\ArrowLine(85,-45)(110,-45)
\ArrowLine(50,-45)(85,-45)
\ArrowLine(-10,-45)(50,-45)

\Text(55,18)[c]{\small $\overline\Psi_{\!R}^d$}

\Text(140,13)[c]{\small $s_L$}
\ArrowLine(110,0)(50,0)

\Text(90,-80)[c]{\small $ \overline\Psi_{\!R}^d$}
\Line(18,3)(24  ,-3)
\Line(24,3)(18,-3)

\Text(155,-78)[c]{\small $\overline b_L$}
\Line(82,-42)(88,-48)
\Line(88,-42)(82,-48)

}
\end{picture}
}
\end{center}
\vspace*{-0cm}
\caption{Leading contribution to
 $B_{s}-\overline B_{s}$ mixing through $\Psi$ and $\Theta$  interactions.
 Other  diagrams differ only by the 
 placement of $\overline \Psi_R Q_L$ mass insertions, and are suppressed by additional powers of $m_{s}$. }
\label{FCNC}
\end{figure}
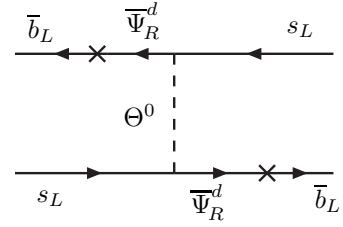

Since $\Psi$ has flavor-dependent couplings, its interactions 
can contribute to flavor-changing neutral processes. The largest couplings 
are to the 3rd and perhaps 2nd generation quarks, so that we expect that the 
most prominent effect is in $B_{s}-\overline B_{s}$ meson mixing. This proceeds 
through the tree-level diagram in  Figure \ref{FCNC}. 
 Integrating out  $\Theta $ and  $\Psi$ generates the effective four-Fermi operator
\be
\!\!\!{\cal L}_{B_s-\overline B_s} = \left( \!   \frac{ C_{23} \, m_{b}}{ 2 M_{\Theta} m_\psi } \! \right)^{\!\!2} \! \! \left(\, \overline b_{R}  
\,T^{a} s_{L}    \right)^{2} + {\rm H.c.} 
\label{mixingop}
\ee
Here we have used the fermion field equations and ignored terms suppressed by factors of $m_{s}/m_{b}$.

 The matrix element of the Hamiltonian due to $\Theta^0$ exchange is
 \be
\hspace*{-0.3cm}
  \left< \overline B_{s} | {\cal H}_\Theta  |B_{s} \right>   \simeq 
\left( \! \frac{ C_{23}  }{ M_\Theta  m_\psi}   \! \right)^{\!\!2}\!\!
M_{B_{s}}^4    f_{B_{s}}^2 
 \etaQCD \,  \frac{5 B_{2} + 3 B_{3}}{288}  ~,
  \hspace*{-1cm}\nonumber \\  
\label{usmix}
\ee
where $M_{ B_{s} } $ and $ f_{B_{s}} = (231\pm15) \,\MeV $ \cite{Gamiz:2009ku}
 are the $B_s$ meson mass and decay constant respectively; $\etaQCD \simeq 1.7$  is the QCD correction  
for the operator in  Eq. \!\!(\ref{mixingop}) due to running from the scale $M_{\Theta}$ down to $M_{B_{s}}$ \cite{Buras:2001ra};  
$B_{2 }\simeq 0.80 $ and $B_{3} \simeq 0.93$ are lattice ``bag'' parameters  \cite{Becirevic:2001xt} for the 
singlet-singlet and octet-octet  color structures arising from 
operator (\ref{mixingop}), respectively.


It is convenient to parametrize the contribution to $B_s$ mixing from $\Theta^0$ relative to the standard model one as
\be
\frac{\left< \overline B_{s} |  {\cal H}_{\rm SM} + {\cal H}_\Theta   |B_{s} \right>}{\left< \overline B_{s} | {\cal H}_{\rm SM}   |B_{s} \right>}
\equiv C_{B_{s}} e^{-i\phi_{s}} ~,
\label{parametrization}
\ee
where $C_{B_{s}}$ is a positive parameter and $-\pi < \phi_{s} < \pi $ is a phase.
The standard model contribution can be extracted from the estimate given in \cite{Lenz:2006hd,Dobrescu:2010rh}:
\be
\left< \overline B_{s} | {\cal H}_{\rm SM}   |B_{s} \right> \approx\! \left(8.0 \!\times\! 10^{-6} \, {\rm GeV}\right)^2 \!(1 \pm 0.15) ~.\nonumber \\
\ee
The 15\% theoretical uncertainty shown above loosens the 
constraint on $C_{B_{s}}$ set by the measured $B_s$ mass difference: $C_{B_{s}} \approx  0.98 \pm 0.15 $.
Comparing Eqs.\! (\ref{usmix}) and (\ref{parametrization}) we find 
\bear
m_\psi &\!\!\!\!=\!\!\!\!&  1.1 \;  {\rm TeV} \times |C_{23}|
\left(\frac{150 \,\GeV}{ \! M_{\Theta}}\right)
\nonumber \\ [2mm]
&& \times \left( C_{B_{s}}^{2}\! +\!1\! - \!2 C_{B_{s}} \!\cos \phi_{s} \right)^{\! - 1/4}  ~,
\label{scalebound}
\eear
and a less illuminating expression of  $\phi_s$ in terms of the phase of $C_{23}$.
For  $M_{\Theta} = 150$ GeV, $C_{23} = 0.2 $, $C_{B_{s}} = 0.9$, and a small CP-violating phase
$\phi_{s} = - 5^\circ$, we get $m_{\psi} = 568$ GeV. However, if the phase is large, as suggested by the 
D0 like-sign dimuon asymmetry \cite{Abazov:2010hv}, then $m_{\psi}$ is below the electroweak scale;
for example $\phi_{s} = - 45^\circ$ gives $m_{\psi} = 260$ GeV. Such a light vector-like quark is not ruled out. 
Note that the main decay mode is likely to be  $\Psi^d \to \Theta^+ \bar{c} \to (c \bar{s})  \bar{c}$, so that 
$\Psi$ pair production leads to a 6-jet final state. The CDF search  \cite{6jet}
in a similar channel gives a lower limit on the $3j$ resonance mass below 200 GeV.

This model also contributes to $b\to s\gamma$ decays at the 1-loop level. Since these
diagrams involve two mass insertions and suffer additional loop suppression, we expect
their contributions to be small. 

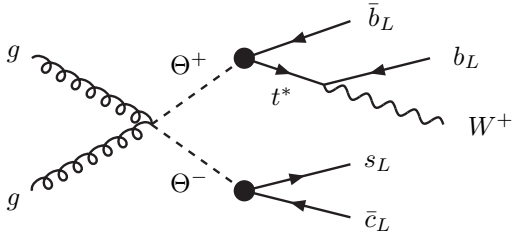
\begin{figure}[t]
\vspace*{0.1cm}
\begin{center}
{
\unitlength=0.7 pt
\SetScale{1.}
\SetWidth{0.9}      
\normalsize    
{} \allowbreak
\begin{picture}(100,100)(70,40)
\Gluon(0,75)(45,50){2.5}{7}
\DashLine(45,50)(80,75){3} \Vertex(80,75){4} 
\ArrowLine(120,89)(80,75)\ArrowLine(80,75)(110,65)
\Photon(155,50)(110,65){-2.}{5} \ArrowLine(150,75)(110,65)
\Gluon(0,25)(45,50){2.5}{7}  
\DashLine(45,50)(80, 25 ){3} \Vertex(80,25){4} 
\ArrowLine(80,25)(120,35)
\ArrowLine(120,15)(80,25) 
\Text(187,19)[c]{\small $\bar c_L$}
\Text(187,51)[c]{\small $s_L$}
\Text(189,130)[c]{\small $\bar{b}_L$}\Text(235,108)[c]{\small $b_L$}
\Text(248,72)[c]{\small $W^+$}\Text(135,87)[c]{\small $t^*$}
\Text(-10,30)[c]{\small $g$}
\Text(-10,110)[c]{\small $g$}
\Text(85,41)[c]{\small $\Theta^-$}
\Text(85,105)[c]{\small $\Theta^+$}
\end{picture}
}
\end{center}
\vspace*{.1 cm}
\caption{Representative diagram for pair production of charged octo-triplets through gluon fusion with a $(jj)(Wb\bar{b})$ final state. 
The $\bullet$ symbol denotes dimension-5 operators induced by the $\Psi$ fermion.
Similar diagrams lead to $4j$, $(W^+b\bar{b})(W^-b\bar{b})$,$(W^+b\bar{s})(jj)$, or $(jb)(Wb\bar{b})$ final states.}
\label{fig:signal}
\end{figure}

\subsection{Dijet resonance plus a $W$ boson at the Tevatron}
\label{sec:Tevatron}

Pair production of octo-triplets, through their QCD couplings to gluons, gives rather large 
cross sections at the Tevatron, as shown in Figure \ref{fig:cross-section}.
In the presence of the vectorlike quark  $\Psi$ and assuming that the trilinear coupling $\mu_\Theta$ is small enough
(see section 3.1), the main decay modes of $\Theta^+$ 
are into a pair of jets ($c\bar{b}$ or $c\bar{s}$) and into $Wb\bar{b}$ ($Wb\bar{s}$ is also possible, but at least 
one $b$ quark is always present due to the decay through the off-shell top quark).
One of the final states (see Figure \,\ref{fig:signal}) arising from $\Theta^+\Theta^-$ 
production is then $(jj)(Wb\bar{b})$, where $j$ is any jet and the parantheses indicate a resonance at $M_\Theta$.
The branching fractions depend on the $|C_{22}|$ and $|C_{33}|$
parameters, and are also quite sensitive to $M_\Theta$, as discussed at the end of section 3.1.
We expect that next-to-leading order QCD corrections to this process, which affect both production and decays, 
increase the rate by a $K$ factor in the $1 - 1.5$ range.

\begin{figure}
\begin{center}%
\vspace{0cm}
\hspace*{-3.2cm}
\parbox{2.2in}{\includegraphics[width=.48 \textwidth]{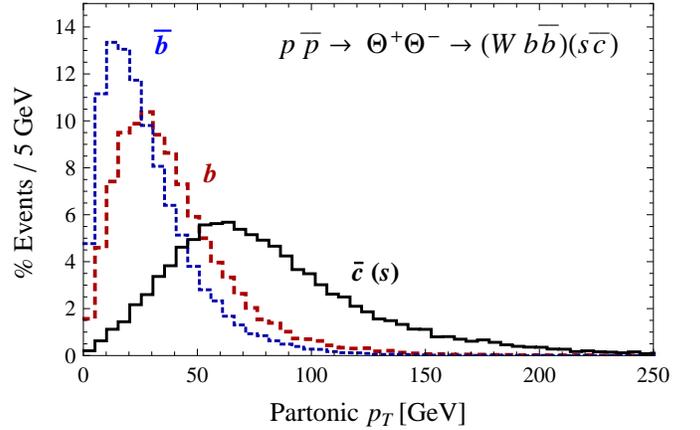}}
\vspace{-0.cm}
\caption{ Partonic $p_{T}$ distributions for the quarks arising from the 
$\Theta^{+} \Theta^{-}\rightarrow (c\bar{s})(W^-b\bar{b})$ process (see Figure \ref{fig:signal}) with $M_{\Theta} = 155$ GeV. 
The $c$ or $\bar{s}$ distribution (black solid line) peaks at higher $p_{T}$ and has a longer tail than the 
$b$ and $\bar{b}$ distributions (blue dotted and red dashed lines). 
\label{fig:pt} }
\end{center}%
\vspace{-0.3cm}
\end{figure}

The two $b$ jets arising from the decay through an off-shell top quark typically have energies below
$(M_\Theta - M_W)/2- m_b$, so are softer than those
arising from the 2-body decay, which typically have energies around $M_\Theta/2$.
Figure \,\ref{fig:pt} shows the transverse momentum of each quark in the $(c\bar{s})(Wb\bar{b})$ final state
for $M_\Theta = 155$ GeV, computed with MadGraph 5 \cite{Alwall:2011uj} with model files generated by FeynRules \cite{Christensen:2008py}.
Given that the quarks from the 2-body decay have the highest $p_T$, the invariant mass distribution 
of the two leading jets from the $p\bar{p}\! \to\! \Theta^+\Theta^-\! \to (jj)(Wb\bar{b})$  process 
exhibits a peak near $M_\Theta$.

\begin{figure}
\vspace{0.02cm}
\begin{center}
\hspace{-3.6cm}
\parbox{2.2in}{\includegraphics[width=.48 \textwidth]{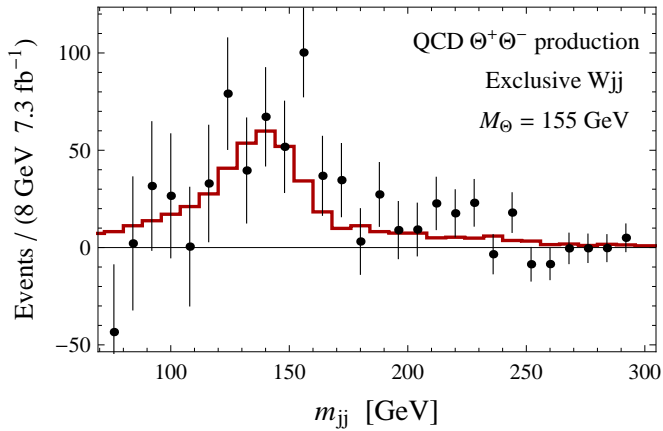}}
\caption{ Invariant mass distribution for the leading two jets arising from the 
$\Theta^{+} \Theta^{-} \rightarrow (jj)(Wb\bar{b}) \rightarrow \ell \nu +4j$ process, where $\ell = e, \mu$,
at the Tevatron with exactly 2 jets passing the cuts.
The red solid line represents events in our simulation 
for ${\mathcal B}( \Theta^\pm\! \rightarrow W^\pm b\bar{b}) = 40\%$ and $M_{\Theta} = 155$ GeV.
The data points with $1 \sigma$ statistical error bars 
are taken from the CDF excess region (Fig. 2 of \cite{CDF7}) 
after the background (including $WW/WZ$) has been subtracted, with the normalization of the 
CDF $Wjj$ background increased by 1\%. 
 \label{fig:main}}
\end{center}
\vspace{-0.6cm}
\end{figure}

In order to compare this signal with the CDF dijet excess \cite{Aaltonen:2011mk}, we generate partonic events using MadGraph 5
for the $p \bar{p} \to \Theta^+\Theta^-\!\!  \to (jj)(Wb \bar{b})$ process with $W\!  \to e\nu, \mu\nu, \tau\nu$.
We then use {\sc Pythia} 6.4 \cite{pythia} for hadronization and parton showering, and 
PGS 4 \cite{pgs} for detector-level effects.
We impose\footnote{We impose cuts on the PGS output using a modified version of the Chameleon package \cite{chameleon}.} the same cuts as CDF \cite{Aaltonen:2011mk}: \
lepton $p_T^\ell > 20$ GeV and $|\eta^\ell| < 1$, 
missing transverse energy $\displaystyle{\not}E_{T} > 25$ GeV,
transverse $W$ mass $M_T(W)>$ 30 GeV,
jet $p_T^j > 30$ GeV and $|\eta^j| < 2.4$,
separation between jets $|\Delta \eta_{jj}| < 2.5$, 
azimuthal separation between  the missing $E_T$ and the leading jet $|\Delta \phi | > 0.4$,
and $p_{Tjj} \ge 40$ GeV for the leading dijet system. 
The resulting dijet invariant mass ($m_{jj}$) distribution for events with exactly 2 jets
is shown in Figure~\ref{fig:main} (solid red line) 
for $M_\Theta = 155$ GeV and a branching fraction ${\cal B}_3 \equiv {\cal B} (\Theta^+\!\!\to W^+b\bar{b}) = 40$\%. 
The rate for this process (before cuts and without including the $W \to \ell\nu$ branching fraction) is
\be
2  {\cal B}_3 (1 - {\cal B}_3) \;  \sigma\!\left( p\bar{p}\to\Theta^{+} \Theta^{-} \right) \simeq 3.2 \; {\rm pb}  ~.
\label{rate}
\ee
The acceptance of the cuts is 6.2\%, so that in 7.3 fb$^{-1}$ of data there are about 470 $Wjj$ events due to $\Theta^+\Theta^-$ production.
The high-mass tail of the $m_{jj}$  distribution is mainly due to events in which the two hardest jets come 
from different octo-triplets.

To compare our simulated $m_{jj}$ with the CDF data shown in Figure 2 (left-side plot) of \cite{CDF7}
we need to subtract all standard model background. 
The CDF Collaboration has fitted the normalization of the large CDF $Wjj$ background to the data 
assuming a Gaussian shape for the signal. 
In the presence of the wider shape arising from our  $\Theta^+ \Theta^-$ signal
the $Wjj$ background normalization is likely to change; increasing it by 1\% gives a reasonable agreement between 
our $m_{jj}$ and the CDF data after background subtraction (Figure~\ref{fig:main}).

If the $K$ factor accounting for the QCD corrections is significantly larger than 1.0, then 
${\cal B}_3$ should be decreased while keeping the rate in Eq.~(\ref{rate}) fixed.
The highest data point, in the $152-160$ GeV bin could indicate that $M_\Theta$ values larger than 155 GeV are preferred. 
However, the jet reconstruction performed by our PGS simulation is likely to be less efficient 
than the CDF reconstruction, so that a larger fraction of the hadrons is missed, reducing the jet energy.
Thus, the dijet mass distribution in Figure ~\ref{fig:main} is likely to be artificially shifted to lower $m_{jj}$
compared to the data, implying that masses even below 155 GeV may be acceptable.
For $M_\Theta = 150$ GeV the cross section is larger by a factor of 1.24,
so that an acceptable fit is obtained for a smaller ${\cal B}_3 \approx 26$\%.

The D0 search \cite{Abazov:2011af} in the same channel with 4.3 fb$^{-1}$ has ruled out a 1.9 pb signal at the 95\% confidence level,
based on the assumptions that the dijet resonance $X$ has a Gaussian shape with a width of 15.7 GeV 
and is produced like the Higgs boson, $p\bar{p} \to W^* \to WX$, through a virtual $W$. 
Clearly, neither of these assumptions applies to our explanation for the CDF excess. The shape of our 
dijet invariant mass distribution is quite different than a Gaussian: it has a high tail below the peak due to 
final state radiation, and it has a long tail above the peak due to the two additional jets from $\Theta$ decay.
The different shape is important because the fit of the background plus signal could improve significantly in the 
presence of our flatter signal shape compared to the pointy Gaussian.
The production through  $\Theta^+\Theta^-$ is also very different than through $WX$, and leads to a different acceptance.
Hence, the D0 result cannot rule out our $Wjj$ signal.

\begin{figure}
\vspace{0.02cm}
\begin{center}
\hspace{-3.6cm}
\parbox{2.2in}{\includegraphics[width=.48 \textwidth]{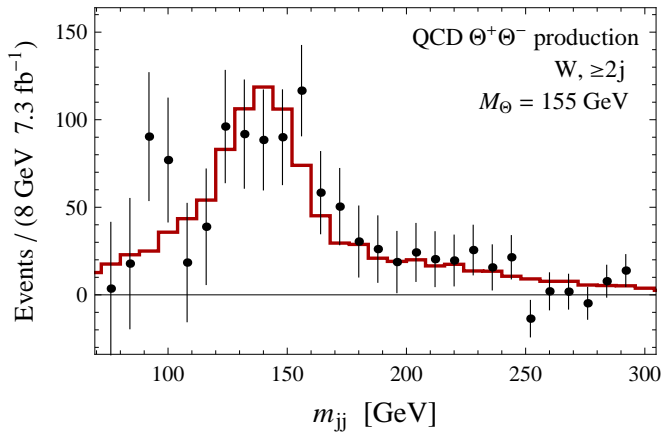}}
\caption{ Same as Figure \ref{fig:main} except that two or more jets pass the cuts. The 
data points with error bars are taken from the CDF excess region (Fig. 5 of \cite{CDF7}) 
after background subtraction with the CDF $Wjj$ background normalization reduced by 3\%. 
 \label{fig:main-incl}}
\end{center}
\vspace{-0.6cm}
\end{figure}

The requirement in the exclusive $Wjj$ search \cite{Aaltonen:2011mk} that exactly two jets pass the cuts 
rejects events arising from $\Theta^+ \Theta^-$ production where one of the $b$ jets has
$p_T^j > 30$ GeV. These events, however, show up in the inclusive $Wjj$ search (Fig. 5 of \cite{CDF7}) 
where two or more jets pass the cuts. 
The normalization of the large $Wjj$ background is fitted to the data 
independently in the exclusive and inclusive cases. 
The additional events mentioned above require the normalization of the 
CDF inclusive $Wjj$ background to be reduced. Figure~\ref{fig:main-incl} shows that 
the QCD production of  $\Theta^+ \Theta^-$ gives a  $W + n$ jet signal with $n\ge 2$ that is consistent
with the CDF data when the normalization of the CDF inclusive $Wjj$ background is reduced by 3\%.

There are a few experimental tests of this interpretation of the CDF excess. 
Even though the two $b$ jets are relatively soft, the fraction of events that 
have a 3rd jet that passes all the cuts is large enough to allow the $b$ tagging of the 
3rd hardest jet. Furthermore, the additional two $b$ jets allow the reconstruction of the full event.  
One complication here is that there is a large background from semileptonic $t\bar{t}$ events. 
Nevertheless, the signal has the property that the reconstructed $W$ boson together with
the two $b$ jets form an invariant mass peak at $M_\Theta$, so that it can be separated from the 
background.

Another test is the process where both octo-triplets decay through an off-shell $t$ quark,
$p \bar{p} \to \Theta^+\Theta^- \to (W^+b\bar{b})(W^-b\bar{b})$. The rate for this is smaller by a 
factor of $2 (1/{\cal B}_3 - 1) \approx 3$ than for the  $(jj)(Wb\bar{b})$ signal. Although this signal 
also suffers from a large $t\bar{t}$ background, it may be observable due to its relatively large rate 
of $\sim 1$ pb at the Tevatron.

Given that the $W$ boson in the  $(jj)(Wb\bar{b})$ signal originates from a decay through an off-shell top quark,
there is no similar signal involving a $Z$ boson or a photon.

The process $\Theta^{+} \Theta^{-} \to (W b \bar{b})(jj)$ 
may affect measurements of the $t\bar t$ cross section in the lepton-plus-jets final state.
However, these measurements typically rely on algorithms trained specifically
to find top pairs and are, thus, less sensitive to new particles that decay into
similar final states. Measurements involving
$b$-tags \cite{Abazov:2008gc} may be sensitive to octo-triplet decays, but their $W$-plus-jets
background normalization is fitted to the data so that they do not necessarily constrain
octo-triplet decays. Furthermore, $b$-tagging efficiency  decreases for softer jets such as our $b$ and $\bar{b}$
(see Figure \ref{fig:pt}).

\begin{figure}[t]
\vspace*{0.1cm}
\begin{center}
{
\unitlength=0.7 pt
\SetScale{1.}
\SetWidth{0.9}      
\normalsize    
{} \allowbreak
\begin{picture}(100,100)(60,40)
\ArrowLine(0,50)(-30,62)\Gluon(0,50)(45,50){2.5}{5}
\DashLine(45,50)(80,75){3} \Vertex(80,75){4} 
\ArrowLine(120,89)(80,75)\ArrowLine(80,75)(110,65)
\Photon(155,50)(110,65){-2.}{5} \ArrowLine(150,75)(110,65)
\ArrowLine(-30,38)(0,50)
\DashLine(45,50)(80, 25 ){3} \Vertex(80,25){4} 
\ArrowLine(80,25)(120,35)
\ArrowLine(120,15)(80,25) 
\Text(187,19)[c]{\small $\bar c_L$}
\Text(187,51)[c]{\small $s_L$}
\Text(189,130)[c]{\small $\bar{b}_L$}\Text(235,108)[c]{\small $b_L$}
\Text(248,72)[c]{\small $W^+$}\Text(135,87)[c]{\small $t^*$}
\Text(-30,45)[c]{\small $q$}\Text(-30,100)[c]{\small $\bar{q}$}
\Text(25,55)[c]{\small $G^\prime$}
\Text(85,41)[c]{\small $\Theta^-$}
\Text(85,105)[c]{\small $\Theta^+$}
\end{picture}
}
\end{center}
\vspace*{.1 cm}
\caption{Same as Figure \ref{fig:signal} except 
the pair of octo-triplets is resonantly produced through an $s$-channel coloron.}
\label{fig:signal-res}
\end{figure}
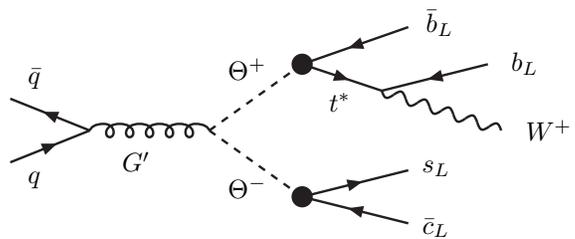

\subsection{Resonant production of $\Theta^+\Theta^-$}
\label{sec:res}

A mass near 150 GeV also appears in another deviation from the standard model:
preliminary CDF data in the $3b$ final state shows an excess in the invariant mass distribution of the leading 
two jets \cite{HiggsPlusb}. That deviation may arise from the 
$\Theta^0\Theta^0 \to (b\bar{b})(b \bar{b})$ process \cite{Bai:2010dj}. 
The transverse energy distributions of the jets in that case appear to favor a pair production mechanism through an $s$-channel
resonance rather than through QCD. The simple renormalizable coloron model presented in \cite{Bai:2010dj}
can be easily adapted to include the octo-triplet discussed here. It is sufficient to charge the scalar field $\Sigma$
(responsible for breaking the $SU(3)\times SU(3)$ extension of the QCD gauge group \cite{Hill:1991at}) under $SU(2)\times U(1)_Y$,
as proposed in \cite{Hall:1985wz}.
The color-octet scalars present in the spectrum can be identified with our 
$\Theta^\pm$ and $\Theta^0$ (although a small mass splitting can be induced by the Higgs VEV), and they couple to the 
coloron field $G^\prime_\mu$ as follows
\bear 
g_s\frac{1\!-\!\tan^2\theta}{2\tan\theta} f^{abc} G^{\prime a}_\mu && \hspace{-.7cm}
\left[ \left( \Theta^{b+} \partial^\mu \Theta^{c-} \!\!+ {\rm H.c.} \!\right) \right.
\nonumber \\ 
&& \hspace{-.7cm} \left. + \, \Theta^{b0} \partial^\mu \Theta^{c0} \right] ~~.
\eear
Here $\tan\theta$ is a parameter in the $0.1 - 0.3$ range. The coloron couples to quarks proportional to 
$g_s\tan\theta$, while it couples only in pairs to gluons at renormalizable level \cite{Dobrescu:2007yp}. 
Thus, single $G^\prime_\mu$ production proceeds entirely through quark-antiquark collisions.

\begin{figure}
\vspace{0.02cm}
\begin{center}
\hspace{-3.6cm}
\parbox{2.2in}{\includegraphics[width=.48 \textwidth]{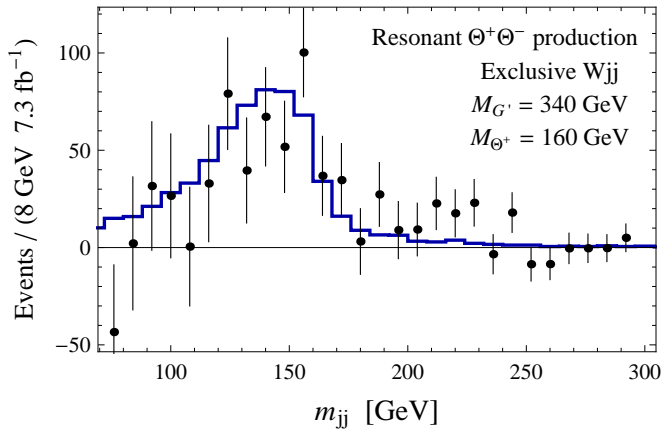}}
\caption{Same as Figure \ref{fig:main} except that a coloron contributes in the  $s$-channel to $\Theta^+\Theta^-$ production. 
The blue solid line represents events in our simulation 
for ${\mathcal B}( \Theta^\pm\! \rightarrow W^\pm b\bar{b}) = 3.9\%$,  $M_{\Theta^+} = 160$ GeV, $M_{G^\prime} = 340$ GeV,
$\tan\theta = 0.15$ and $\Gamma_{G^\prime} = 6.5$ GeV.  
 \label{fig:coloron-excl}}
\end{center}
\vspace{-0.5cm}
\end{figure}

The resonant  $p\bar{p}\to G^\prime_\mu \to \Theta^+\Theta^-$ production (Figure \ref{fig:signal-res}) may be an order of 
magnitude larger than QCD pair production  \cite{Bai:2010dj}.
This theory preserves the good  agreement with the CDF data shown in Figure~\ref{fig:main} provided the branching fraction 
${\cal B}_3$ is decreased accordingly.  
The width of the coloron $\Gamma_{G^\prime}$ is sensitive to $\tan\theta$ and to the octet masses. 
Assuming that the coloron decays only 
into $q\bar{q}$, $\Theta^+\Theta^-$ and $\Theta^0\Theta^0$ (additional decay channels may increase the width \cite{Bai:2010dj}),
we find  $\Gamma_{G^\prime} $  in the $3.2 - 6.5$ GeV range for $M_{\Theta^+} = 160$ GeV, $\tan\theta = 0.15$,  
a coloron mass $M_{G^\prime} = 340$ GeV, and  $M_{\Theta^0} $ in the $160-140$ GeV range.
For this set of parameters with $\Gamma_{G^\prime} = 6.5$ GeV we generate events as described in section 3.3. 
The invariant mass distribution of the leading two jets is shown in Figure \ref{fig:coloron-excl} for 
a rate 
\be
\! \!\! \!\! \!2  {\cal B}_3 (1 - {\cal B}_3) \;  \sigma\!\left( p\bar{p}\to G^\prime\! \to \Theta^{+} \Theta^{-}
\right) \simeq 3.8 \; {\rm pb} ~.
\label{rate-resTeV}
\ee
Acceptance (without including $W$ branching fractions) is 7.3\% for this process. The branching fraction inferred from the above rate is small, ${\cal B}_3 = 3.9\%$,
implying that the coloron $\Theta^+ \Theta^-$ production  
dominates by an order of magnitude over the QCD $\Theta^+ \Theta^-$ contribution. Nevertheless, we include in Figure \ref{fig:coloron-excl}
both production mechanisms and their interference, as well as the electroweak $\Theta^+ \Theta^-$ production.

\begin{figure}
\vspace{0.02cm}
\begin{center}
\hspace{-3.6cm}
\parbox{2.2in}{\includegraphics[width=.48 \textwidth]{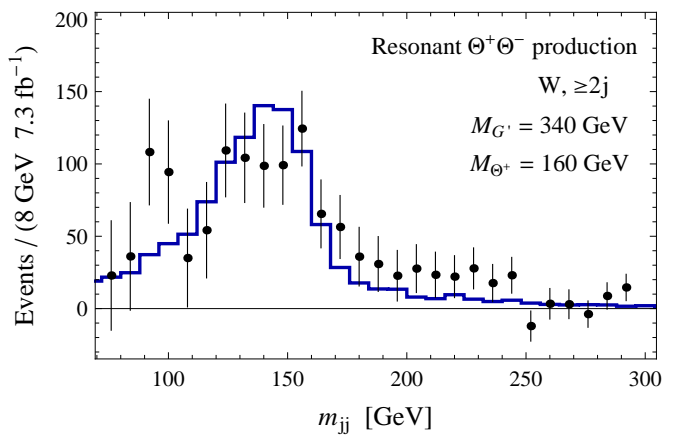}}
\caption{ Same as Figure \ref{fig:coloron-excl} except that two or more jets pass the cuts. The 
CDF data points are taken from Fig. 5 of \cite{CDF7} 
after background subtraction with the normalization of the CDF $Wjj$ background reduced by 5\%. 
 \label{fig:coloron-incl}}
\end{center}
\vspace{-0.6cm}
\end{figure}

Figure \ref{fig:coloron-incl} shows the $m_{jj}$ distribution in the inclusive case ($W$ plus two or more jets),
with the normalization of the CDF $Wjj$ background reduced by 5\%.
The subtracted data is consistently higher than the signal in the $m_{jj} \approx 170 -240$ GeV range, so one could 
conclude that the QCD production mechanism (see Figure \ref{fig:main-incl}) provides a better description of the CDF data.
However, next-to-leading order effects are not included in these figures, and it is conceivable that they 
sufficiently raise the high-mass tail of the resonant production shown in Figure \ref{fig:coloron-incl}.
Furthermore, the CDF result for the inclusive case (Fig. 5 of \cite{CDF7}) does not include systematic errors.
We also emphasize that our detector simulation using PGS 4 \cite{pgs} is only a rough approximation to the CDF full detector simulation.

\begin{figure}[t]
\begin{center}
\hspace{-3.6cm}
\parbox{2.2in}{\includegraphics[width=.48 \textwidth]{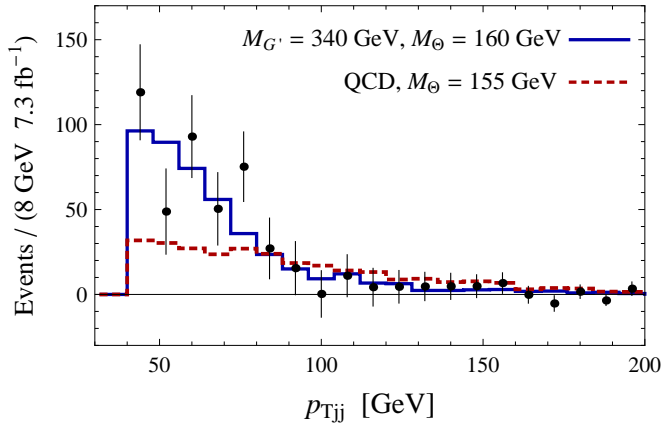}}
\caption{ $p_T$ distribution of the dijet system for events satisfying $115$ GeV $ \leq m_{jj} \leq 175$ GeV. 
The blue solid line is for a coloron with the same parameters as in Figure \ref{fig:coloron-excl},
while the red dashed line is for QCD $\Theta^+\Theta^-$ production with $M_{\Theta^+} = 155$ GeV.
The CDF data points with statistical error bars are taken from Fig. K9 of \cite{CDF7kin}
after background subtraction consistent with Fig. \ref{fig:main}. 
 \label{fig:pTjj}}
\end{center}
\vspace{-0.02cm}
\end{figure}

\begin{figure}[h!]
\begin{center}
\hspace{-3.6cm}
\parbox{2.2in}{\includegraphics[width=.48 \textwidth]{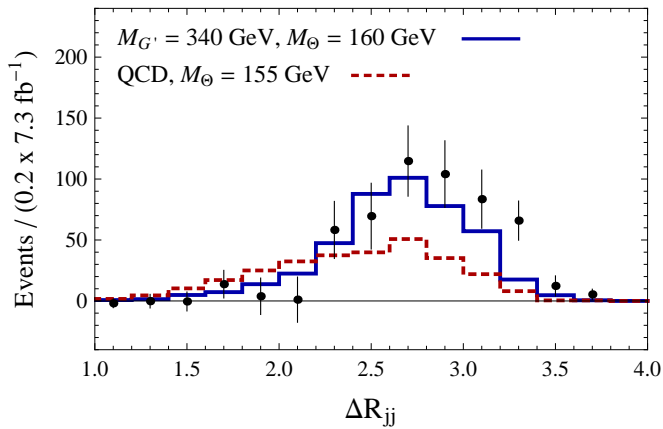}}
\caption{ Same as Figure \ref{fig:pTjj} for the angular separation $\Delta R_{jj}$ of the two leading jets
(CDF data taken from Fig. K7 of \cite{CDF7kin}).
 \label{fig:DeltaR}}
\end{center}
\vspace{-0.02cm}
\end{figure}

A better discriminant between the resonant and QCD production mechanisms is provided by the CDF kinematic distributions  \cite{CDF7kin}
 for the exclusive search in the  $m_{jj} \approx 115 - 175$ GeV window.
The transverse momentum distribution of the dijet system (Figure \ref{fig:pTjj}) shows that resonant production 
fits the data much better than QCD  $\Theta^+ \Theta^-$ production. We reach the same conclusion using the $\Delta R_{jj}$ 
distribution of the angular separation between the two jets (Figure \ref{fig:DeltaR}).
Although some of the data points are not well fitted (the $p_{Tjj} =72-80$ GeV bin and the $\Delta R_{jj} = 3.2-3.4$ bin) by our theoretical 
predictions, the shapes of both the $p_{Tjj}$  and $\Delta R_{jj}$ distributions are in remarkable agreement.  
In both Figures  \ref{fig:pTjj} and  \ref{fig:DeltaR} we use the same background subtraction as in Figure \ref{fig:main}, where only one background 
($Wjj$ with combined electron and muon contributions) is rescaled (increased by 1\%).
We expect that a fit of the standard model background plus our signal, 
where various background normalizations are allowed to vary, would improve the agreement between  $\Theta^+\Theta^-$
production and the CDF $Wjj$ excess.

\par\medskip

\subsection{LHC Signals}
\label{sec:LHC}

QCD $\Theta^+\Theta^-$ production, which proceeds through gluon-gluon collisions, 
 is two orders of magnitude  larger at the 7 TeV  LHC than at the Tevatron 
(see Figure \ref{fig:cross-section}), so that the $(jj)(Wb\bar{b})$ signal discussed in section 3.3 
will soon be within the reach of the CMS and ATLAS experiments. 
Using typical parameters that explain the CDF dijet resonance, 
$\mathcal B_3 = 40\%$  
 and $M_{\Theta} = 155$ GeV,  we find
that the process in Figure \ref{fig:signal} has a leading-order rate (before cuts) of
\be
\sigma\!\left( pp\to\Theta^{+} \Theta^{-} \!\!\to  (jj) (\ell\nu b\bar{b}) \right) \simeq 52 \; {\rm pb} ~~, 
\label{LHCrate}
\ee
where $\ell = e,\mu$.
Furthermore, the $(W^+b\bar{b})(W^-b\bar{b})$ process also has a large rate, suppressed 
only by a factor of $2(1/B_3 -1) \approx 3$ compared to $(jj)(Wb\bar{b})$, 
so the fully leptonic $(\ell^+\nu b\bar{b})(\ell^-\bar{\nu}b\bar{b})$ signal has a cross section of 3.7 pb
and will also be soon within the reach of the LHC.

In the non-minimal model (section 3.4 and Figure \ref{fig:signal-res}) where
resonant $G^\prime_\mu \to \Theta^{+} \Theta^{-}$ production is the main process responsible
for the CDF excess, the $(jj) (\ell\nu b\bar{b})$ rate at the  7 TeV LHC is reduced by a factor of $\sim 5$ 
compared to Eq.~(\ref{LHCrate}), due to smaller 
parton distributions for quark-antiquark collissions:
\be
\sigma\!\left( pp\to G^\prime\! \to \Theta^{+} \Theta^{-} \!\!\to  (jj) (\ell\nu b\bar{b}) \right) \simeq 10 \; {\rm pb}~ . \;\;\;
\label{LHCrate-res}
\ee
Although QCD $\Theta^+\Theta^-$ production is still present in the coloron model, explaining the CDF signal
requires a 10 times smaller ${\cal B}_{3}$ branching fraction, which reduces the gluon initiated contribution to the $(jj)(Wb\bar{b})$ signal. The 
smaller  ${\cal B}_{3} \approx 3.9\%$ also suppresses the $(W^+b\bar{b})(W^-b\bar{b})$ signal in this model (the rate is 43 fb).

While our analysis has emphasized the region around 
 $M_{\Theta} = 150$ GeV, future searches could discover much heavier octo-triplets
 which decay into final states involving top quarks. For non-negligible values of 
the $C_{33}$ parameter, the processes $\Theta^{+}\Theta^- \to (t\bar{b})(\bar{t}b)$ and
$\Theta^0\Theta^0\to 4t$ are important tests of the octo-triplet decaying through higher-dimensional 
operators (these final states have been studied in \cite{Gerbush:2007fe}).


\section{Conclusions}\setcounter{equation}{0}

We have shown that the renormalizable extension of the standard model with one octo-triplet
({\it i.e.}, a scalar in the adjoint representation of the standard model gauge group)
involves two new parameters: the octo-triplet mass $M_{\Theta}$ and cubic self-coupling 
$\mu_{\Theta}$. 
For $\mu_{\Theta}^2/M_{\Theta} \gg 10^{-9}$ GeV the charged octo-triplet almost always decays
into $Wg$ in the absence of other new particles. 
The rate for this 1-loop process is accidentally suppressed (see Appendix C), but the decay is prompt 
as long as  $\mu_{\Theta}^2/M_{\Theta} \gtrsim 10^{-7}$ GeV. 
The neutral octo-triplet decays to $Z g$ or $\gamma g$, with widths comparable to that for 
$\Theta^\pm \!\to\! W^\pm g$.
For $\mu_{\Theta} \to 0$, the main decay is a tree-level 3-body process, $\Theta^\pm \!\to\! \Theta^{0} e^\pm \nu $,
with a displaced vertex, while $\Theta^{0}$ is stable.

At the Tevatron and the LHC, octo-triplets are produced in pairs with relatively large cross sections
(see Figure 2). The main signatures are 
\bear
&& \Theta^+\Theta^- \to (W^+ g)(W^- g) ~~,
 \\
&& \Theta^0\Theta^0 \to (Z g)(Z g) \; , \; (Z g)(\gamma g) \; , \;  (\gamma g)(\gamma g) ~~. \nonumber
\eear
The rates for these processes suggest that Tevatron experiments 
can be sensitive to $M_{\Theta}$ up to a few hundred GeV, and LHC experiments 
to $M_{\Theta}$ above 1 TeV; however, more precise sensitivity estimates require 
detailed studies of the backgrounds.

Since  octo-triplets have very small widths, decays through higher-dimensional operators 
may compete with the 1-loop processes. Operators of the type $\Theta\overline{Q}\slash\!\!\!\partial Q  $
may be induced by a heavy vectorlike quark, and lead to the $\Theta^+ \to t\bar{b}$ decay for 
$M_{\Theta} \gtrsim 175$ GeV. For a lighter  octo-triplet, there is competition between 
the 2-body decays $\Theta^+\!\!  \to c\bar{s}$ or $c\bar{b}$ and the 3-body decays 
$\Theta^+\!\!  \to\!  t^*\bar{b}\!  \to \! Wb\bar{b}$ or  $t^*\bar{s}\!  \to \! Wb\bar{s}$   through an off-shell top quark. 
The neutral octo-triplet decays mainly to $b\bar{b}$, $c\bar{c}$, $W^+b\bar{c}$ and $W^-\bar{b} c$ 
for $M_{\Theta} \lesssim m_t + m_c$, to $t \bar{c}$ and $\bar{t} c$ for larger masses below $2m_t$, 
and to $t\bar{t}$ for masses above $2m_t$.
For a range of parameters, the branching fractions for these decays are larger than the 
ones into a gluon plus an electroweak boson mentioned above. 
The collider signatures for $M_{\Theta} \lesssim 175$ GeV then include 
\bear 
&& \hspace*{-0.71cm}
\Theta^+\Theta^-\! \to (jc)(Wb\bar{b}) \; , \, (bc)(Wb\bar{b}) \; , \,  (W^+b\bar{b})(W^-b\bar{b})  ~,
\nonumber  \\
&& \hspace*{.91cm}
(jc)(j\bar{c}) \; , \; (jb)(j\bar{b})  \; , \; ... \, ,
\nonumber  \\
&& \hspace*{-0.71cm}
\Theta^0\Theta^0 \to (b\bar{b})(b\bar{b}) \; , \; (b\bar{b})(c\bar{c}) \; , \; (b\bar{b})(W bc)  \; , \; ... \, ,
\eear
For a heavier octo-triplet the signatures are mainly $(t\bar{b})(\bar{t}b)$ and 
$(t\bar{c})(\bar{t}c)$, while above 350 GeV the $4t$ final state also opens up.

Signatures of pair production followed by one octo-triplet decaying  
through higher-dimensional operators and the other decaying into a gluon and an electroweak gauge boson at one loop 
are also possible. These include 
$\Theta^+\Theta^-\! \!\to (W g)(j c)$, $\Theta^0\Theta^0\! \to (Z g)(c \bar{c})$ or
$(\gamma g)(c \bar{c})$
and similar processes involving $b$ quarks (or $t$ quarks if kinematically allowed).

Some of the final states mentioned above, namely $(jc)(Wb\bar{b})$, $(jc)(Wbj)$, $(j c)(W g)$, 
may be relevant to the CDF excess \cite{Aaltonen:2011mk}
in the dijet resonance plus $W$ search. 
In the case where $\Theta^+$ decays mostly into $c\bar{s}$ and $W^+b\bar{b}$,
so that the process is $p\bar{p} \to \Theta^+\Theta^- \to (cj)(Wb\bar{b})$, we have 
shown that  the $b$ jets are substantially softer than the jets originating 
from the $\Theta^+\to c\bar{s}$ decay. Events where these $b$ jets do not pass the CDF cuts 
could explain the dijet resonance plus $W$ signal if $\Theta$ has a mass in the $150-170$ GeV
range.

We have compared two production mechanisms of charged octo-triplet pairs: through the 
QCD couplings to gluons (these are always present due to gauge invariance), and through 
an $s$-channel resonance (we have focused on a 
coloron, but a $Z^\prime$ coupled to octo-triplets would not be very different). 
Both mechanisms are consistent with the CDF excess in the dijet invariant mass distribution when exactly two jets are required to pass the cuts.
In the inclusive case (two or more jets pass the cuts), QCD $\Theta^+\Theta^-$ production  fits the CDF data more precisely than resonant production. 
However, this difference is not conclusive 
given that the low tail of the resonant production compared to the background-subtracted data may 
be due to systematic errors in the standard model background, and may also be corrected by a 
fit of the background (with several free normalizations, as usual) plus coloron signal.
Other kinematic distributions obtained by CDF \cite{CDF7kin} can differentiate various models more effectively.
We have shown that the shapes of the transverse momentum distribution for the dijet system ($p_{Tjj}$)
and of the angular separation distribution for the two leading jets  ($\Delta R_{jj}$)
agree rather well with the resonant mechanism while being quite different than the predictions of QCD 
$\Theta^+ \Theta^-$ production.

It is intriguing that almost the same mass ($\sim 150$ GeV) appears in another deviation from the standard model, 
namely the $3b$ CDF search \cite{HiggsPlusb}, which could be attributed to the 
$\Theta^0\Theta^0 \to (b\bar{b})(b \bar{b})$ process \cite{Bai:2010dj}. Resonant production through a coloron 
also agrees better to various kinematic distributions in that case.

The interpretation of the dijet plus $W$ signal in terms of an octo-triplet decaying via an off-shell top quark 
can be tested by the $b$-tagging of the third jet, or by the reconstruction of the 
$Wb\bar{b}$ peak at the same mass as the dijet peak.

At the 7 TeV LHC, if octo-triplet production is  through the 
QCD couplings to gluons, then the dijet-plus-$W$ signal has a large cross section  (52 pb for $M_{\Theta^+} = 155$ GeV)
because it is dominated by gluon fusion. 
In the case of dominant production through  an $s$-channel resonance coupled to $q\bar{q}$
like the coloron, the LHC signal is reduced to $\sim 10$ pb. 

If the couplings of the vectorlike quark are complex, then tree-level $\Theta^0$ exchange induces 
CP violation in $B_s - \bar{B_s}$ mixing. For a  vectorlike quark mass of a few hundred GeV 
(which is allowed because its main decay is into three jets),
this effect can be large enough to produce a significant part of the 
like-sign dimuon asymmetry observed by the D0 Collaboration \cite{Abazov:2010hv}.

We note that similar final states can also arise from a fermiophobic
octo-doublet field. 
However, this has nontrivial couplings to the standard model Higgs doublet, so
the mass splitting between  charged and neutral components may be large.
By contrast, the tiny mass splitting between charged and neutral octo-triplets suppresses the tree-level
3-body decays.
Furthermore, the neutral octo-doublet decays into gluons, while $SU(2)_W$ symmetry forbids 
pure gluonic decays of the neutral  octo-triplet.

\bigskip

\noindent {\bf Acknowledgments:} 
We would like to thank Johan Alwall,  Alberto Annovi, Yang Bai, William Bardeen, Chris Bouchard, Patrick Fox, Walter Giele, Roni Harnik,
David E. Kaplan, Joachim Kopp,  Adam Martin, Olivier Mattelaer, 
 Yuh\-sin Tsai, and Ciaran Williams
for helpful discussions and comments. GZK is supported by a Fermilab Fellowship in Theoretical Physics.
Fermilab is operated by Fermi Research Alliance, LLC, under Contract DE-AC02-07-CH11359
with the US Department of Energy.

\section*{Appendix A: Feynman rules}
\label{sec:rules}
\renewcommand{\theequation}{A.\arabic{equation}}
\setcounter{equation}{0}

The Feynman rules for  octo-triplets, derived 
from Eqs. (\ref{kinetic})-(\ref{triplet-gluons}), are given by:  


\vspace*{0.8cm}

\begin{center} 
\unitlength=0.45 pt
\SetScale{0.45}
\SetWidth{2.}      
\normalsize    
{} \allowbreak
\thicklines


\begin{picture}(100,100)(20,-100)
\put(-148,-0){\vector(1,0){10} }
\put(-81,28){\vector(-1,-1){10} }
\put(-180,-30){
\DashLine(-0,30)(70,30){4} 
\DashLine(70,30)(120,80){4}
\DashLine(70,30)(120,-20){4}
\Text(20,10)[c]{\small $\Theta^{+}_{a}$}
\Text(95,85)[c]{\small $\Theta^{- }_{b}$}\Text(95,-29)[c]{\small $\Theta^{0 }_{c}$}  
\Text(180, 30)[c]{$ = 2 \mu_{\Theta} f^{abc}$}
}
\end{picture}

\begin{picture}(100,100)(-87,-60)
\put(20,65){\vector(-1,-1){25} } \Text(5,68)[c]{\small $p$}
\put(20, -25){\vector(-1,1){25} }\Text(5,-35)[c]{\small $q$}
\put(-80,-10){
\Gluon(15,30)(70,30){-4}{4}                 
\DashLine(70,30)(120,80){4} 
\DashLine(70,30)(120,-25){4}
\Text(15,49)[c]{\small $ G^{a}_{\mu} $}
\Text(115,100)[c]{\small $\Theta^{b}$}\Text(115,-40)[c]{\small $\Theta^{c }$}  
\Text(220, 30)[c]{$ = - g_{s} f^{abc} (p - q)_{\mu}$}
}
\end{picture}

\begin{picture}(100,140)(20,40)
\put(-185, 60){
\Gluon(15,-27)(70,30){-4}{5} \Text(8,-50)[c]{\small$ G_\nu^a$}               
\Photon(70,30)(15, 80){-4}{5}\Text(10,100)[c]{\small $W_\mu^{-}$}  
\DashLine(70,30)(125,85){4}  \Text(106,98)[c]{\small $\Theta^{0}_{b}$}       
\DashLine(70,30)(125,-25){4}\Text(115,-45)[c]{\small $\Theta^{+}_{c}$}
\Text(210,30)[c]{$ = - 2 g_{s}g  f^{abc} g_{\mu\nu}$}  \put(105,-5){\vector(-1,1){10} }
}
\end{picture}

\begin{picture}(100,100)(-106,180)
\put(27,240){\vector(-1,-1){25} } \Text(10,240)[c]{\small $p$}
\put(23, 145){\vector(-1,1){25} }\Text(08,140)[c]{\small $q$}
\put(-75,160){
\Photon(10,30)(70,30){-4}{4}                 
\DashLine(70,30)(115,75){4}  \put(102,-2){\vector(-1,1){10} }
\DashLine(70,30)(115,-15){4}
\Text(26,52)[c]{\small $ W^{-}_{\mu} $}
\Text(125,100)[c]{\small $\Theta^{0}_{a}$}\Text(125,-35)[c]{\small $\Theta^{+}_{b}$}  
\Text(210, 30)[c]{$ = i g\, \delta^{ab} (p - q)_{\mu}$}
}
\end{picture}

\begin{picture}(100,100)(20,70)
\put(-185, 35){
\Gluon(15,-27)(70,30){-4}{5} \Text(8,-50)[c]{\small$ G^{a}_\nu $}               
\Photon(70,30)(15, 80){-4}{5}\Text(10,96)[c]{\small $Z_\mu$}  
\DashLine(70,30)(125,85){4}  \Text(106,98)[c]{\small $\Theta^{+}_{b}$}       
\DashLine(70,30)(125,-25){4}\Text(115,-45)[c]{\small $\Theta^{-}_{c}$}
\Text(240,30)[c]{$ =  -2 g_{s}g \,\cos\theta_{W} f^{abc} g_{\mu\nu}$}  \put(105,-5){\vector(-1,1){10} }
 \put(105,65){\vector(-1,-1){10} }
}
\end{picture}

\begin{picture}(100,100)(-40,200)
\put(29,190){\vector(-1,-1){25} } \Text(10,190)[c]{\small $p$}
\put(26, 92){\vector(-1,1){25} }\Text(10,88)[c]{\small $q$}
\put(-75,110){
\Photon(10,30)(70,30){-4}{4}                 
\DashLine(70,30)(115,75){4}  \put(102,-2){\vector(-1,1){10} }
\DashLine(70,30)(115,-15){4}  \put(102,62){\vector(-1,-1){10} }
\Text(20,52)[c]{\small $ Z^{\mu} $}
\Text(125,100)[c]{\small $\Theta^{+}_{a}$}\Text(125,-35)[c]{\small $\Theta^{-}_{b}$}  
\Text(238, 30)[c]{$ =  i g \cos\theta_{W} \, \delta^{ab} (p - q)_{\mu}$}
}
\end{picture}
\end{center}

\vspace*{2.1cm}

The Feynman rules involving photons are  
identical to those involving a $Z$ boson shown above but with the replacement $g \cos\theta_{W} \to e$. 


\section*{Appendix B: Tree-level 
octo-triplet decay}
\renewcommand{\theequation}{B.\arabic{equation}}
\setcounter{equation}{0}

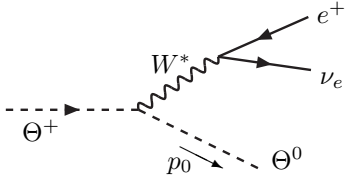
\begin{figure}[t]
\vspace*{-1.cm}
\begin{center}
{
\unitlength=0.7 pt
\SetScale{1.}
\SetWidth{0.9}      
\normalsize    
{} \allowbreak
\begin{picture}(100,100)(40,60)    
\put(-20,0){     
\DashArrowLine(25,35)(75, 35 ){3} 
\DashLine(75, 35)(120,13){3} 
\Photon(75,35)(105,55){2}{6}
\Text(125,75)[c]{\small $W^*$}
\ArrowLine(139,70)(105,55)
\ArrowLine(105,55)(140,50)
\Text(55,39)[c]{\small $\Theta^+$}
\Text(212,103)[c]{\small $e^{+}$}
\Text(212,66)[c]{\small $\nu_{e}$}
\Text(188,24)[c]{\small $\Theta^0$}
\put(130,30){\vector(2,-1){23}}\Text(130,20)[c]{\small $p_0$}
}
\end{picture}
}
\end{center}
\vspace*{0.5cm}
\caption{ \label{fig:lastfig}3-body decay of the charged octo-triplet scalar through an off-shell $W$ boson.}
\end{figure}

In this appendix we compute width for 
the 3-body decay of the charged octo-triplet through an off-shell $W$, as shown in Figure \ref{fig:lastfig}. 
We define $p_{e}, p_{\nu},$ and $ p_{0}$ to be the outgoing momenta for $e, \nu$ and $\Theta^{0}$ respectively.
Using the Feynman rule from Appendix A, the amplitude for this process is 
\be
{\cal M} \simeq \frac{ \sqrt{2}\,g^{2} }{M_{W}^{2} - 2 p_e\! \cdot\!  p_{\nu} } \,\overline{u}(p_{e})\,  \displaystyle{\not} p_{0} P_{L} v(p_{\nu}) ~~,
\ee
where we have used the $e, \nu$ equations of motion in the massless lepton limit. 
Squaring the amplitude and summing over helicities we find
\be 
\hspace*{-1cm}
\overline{|{\cal M}|^{2}}  \!\!\!\!\!  &=& \!\!\!\! 
4 g^{4}\; \frac{2 (p_{e}\! \cdot\!  p_{0})(p_{\nu}\! \cdot\!  p_{0}) - p_{e}\! \cdot\!  p_{\nu} M_{\Theta^{0}}^{2}}{(M_{W}^{2} - 2 p_e\! \cdot\!  p_{\nu})^2}  ~~.
\label{ampsq} 
\ee

The decay width in the $\Theta^{+}$ rest frame is then given by
\bear
&& \hspace*{-1.6cm} \Gamma(\Theta^{+}\!\!\to\! \Theta^{0}e^+\nu)=  \frac{g^{4} M_{\Theta^+}}{16\pi^{3}}
\!\int_{0}^{\varepsilon} 
\!\!dE_{\nu} \! \int_{\varepsilon- E_\nu}^{E_e^{\rm max}} \!\!\! dE_{e}
\nonumber \\ [1.5mm]
&& \hspace*{0.2cm} \times \, \frac{M_{\Theta^+}\left(\varepsilon \! - E_{\nu} - E_{e}  \right) + 2 E_{\nu} E_{e}}
{\left[ M_W^2 - 2M_{\Theta^+}\left( \varepsilon \! - E_{\nu} - E_{e}  \right) \right]^2}  ~, 
\label{integration}
\eear
where $\varepsilon$ is the maximum lepton energy,
\be
\varepsilon 
= \frac{M_{\Theta^{+}}^{2} \!- M_{\Theta^{0}}^{2}}{\,\,2 M_{\Theta^{+}}} ~~, 
\ee
and $E_e^{\rm max}$ is the maximum positron energy for a fixed neutrino energy $E_\nu$,
\be
E_e^{\rm max} = \frac{\varepsilon\! - E_\nu}{ 1 - 2E_{\nu}/M_{\Theta^+}}  ~~.
\ee

The integrals in Eq.~(\ref{integration}) can be performed analytically, with the result 
\bear
&&  \hspace*{-.7cm} 
\Gamma(\Theta^{+}\!\!\to\! \Theta^{0}e^+\nu)=  \frac{\alpha^2 M_W^4}{16\pi \sin^4\!\theta_W  M_{\Theta^+}^3\!\!}
\, {\cal G}\! \left(\!\frac{\varepsilon}{M_{\Theta^+}\!\!} \, ,\frac{\! 2M_{\Theta^+}^2\!}{M_W^2}\!\right) \nonumber \\ [-.5mm]
&& \hspace*{-0.2cm} 
\label{exact}
\eear

\vspace*{-0.4cm} 

\noindent 
where we introduced a function
\bear
&&  \hspace*{-0.8cm} {\cal G}\! \left(x,r\right) = - \xi \left( 1 \!+ \! r  \! - \! r x \right) \ln \left(1 \!-\! x \!+\! r x^2 \!+\! \xi x \right)
\nonumber \\ [1.5mm]
&& \hspace*{0.3cm} 
-\, \frac{1}{2} \left[ \! \left(\frac{1}{2} \! -  \! x \!\right) \!r^2 \!+\! \xi^2 
\!-\! \xi\left( 1 \!+ \! r  \! - \! r x  \!\right) \right] \ln (1  \!- \!2 x)
\nonumber \\ [1.5mm]
&& \hspace*{0.3cm} 
+ \, \frac{r^3 x^3}{3} \!+\! \frac{3}{2} r^2 x \left(1\!-\! x \right) \!+\! r x ~~,
\eear
with 
\be 
\xi \equiv \xi \left(x,r\right) = \left[ 2 r + ( 1 - r x )^2 \rule{0mm}{3.7mm} \right]^{\!1/2} ~~. 
\ee
Interestingly, the expansion of $r^{-4}{\cal G}\! \left(x,r\right)$ for $|x| < 1$ starts at $x^5$, and the leading $r$-dependent 
term arises even later, at $x^7$:
\be
\hspace*{-0.3cm} 
{\cal G}\! \left(x,r\right) = \frac{r^4 x^5\! }{15} \left[ 1 + x +\frac{2}{7} \left(4\!-\!r \right)x^2 \right] \! + \! O\!\left(x^8\right) .
\ee
Translating this expansion into a power series in $\delta M \equiv M_{\Theta^{+}} -M_{\Theta^{0}}$ we find 
that the exact tree-level width of Eq.~(\ref{exact}) is given, up to corrections of order $(\delta M)^8$, by
\bear
&& \hspace*{-1.2cm}
\Gamma(\Theta^{+}\!\!\to\! \Theta^{0}e^+\nu) \simeq \frac{\alpha^2 \, (\delta M)^5}{15\pi \sin^4\!\theta_W  M_W^4\!\!} \,
\left[ 1 -\, 
\frac{\!3(\delta M)\!}{\,2 M_{\Theta^+}\!}\rule{0mm}{4mm} \right.
\nonumber \\ [1.5mm]
&& \hspace*{1.1cm} \left. +\, \frac{4}{7}\! \left(\frac{9}{8} - \frac{M_{\Theta^+\!}^2}{M_W^2}\right) \!\!
\left(\frac{\!\delta M\!}{\!M_{\Theta^+}\!}\!\right)^{\!\!2}\right] . 
\eear

%
%

\section*{Appendix C:  One-loop decay of a scalar octet into gauge bosons}
\label{sec:1loop}
\renewcommand{\theequation}{C.\arabic{equation}}
\setcounter{equation}{0}

In this Appendix we compute the width for a color-octet scalar
 decaying to a gluon and a (massive or massless) vector boson $V_{\mu}$, which 
proceeds through scalar 1-loop diagrams like those of Figure \ref{fig:triangle}. In 
particular, this computation applies to the process  
$\Theta^{\pm} \to W^{\pm}g, \> \Theta^{0}  \to  \gamma g$ or $Z g$.  
\par 
We label the $V_{\mu}$ and  gluon 4-momenta (polarizations) by $p_1$ ($\epsilon_{1}$) and $p_2$ $(\epsilon_{2})$, 
respectively.  Since the gluon is always transversely polarized $( \epsilon_{2}\cdot p_{2}=0)$,  
angular momentum conservation
demands that the other vector also be transverse, so $\epsilon_{1} \cdot p_{1}=0$.
Given that the contraction
 $\epsilon_{\mu\nu\rho\sigma}\, p_{1}^{\mu}\epsilon_{1}^{\nu}\, p_{2}^{\rho}\, \epsilon_{2}^{\sigma}$,
 cannot arise from scalar triangle diagrams,
 the amplitude contains only two terms: $\epsilon_{1}\! \cdot \epsilon_{2} $ 
 and $\left(\epsilon_{1}\!\cdot p_{2} \right) (\epsilon_{2}\!\cdot p_{1}) $. 
 Furthermore, by the Ward-Takahashi identity 
 the amplitude vanishes
 upon replacing $\epsilon_{2}$ with $p_{2}$, so that the most general
 amplitude due to scalar loops is given by  
\be
\mathcal{M} =\!\frac{\mu_{\Theta} g_{s} \tilde g}{\pi^{2}}  \, {\cal C}  \left( \frac{1}{2}\,\epsilon_{1}\! \cdot \epsilon_{2} 
- \frac{ \left( \epsilon_{1}\!\cdot p_{2} \right) \epsilon_{2}\!\cdot p_{1} }{M_{\Theta}^{2} - M_{V}^{2}} \right) .
\label{amplitude2}
\ee
where  $\mu_{\Theta}$ is the scalar trilinear coupling, $\tilde g$ is the scalar-vector gauge coupling, 
$M_{V}$ and $M_{\Theta}$ are the $V_{\mu}$ and scalar masses. 
The dimensionless coefficient $\cal C$ is the only quantity that needs to be computed from loop
integrals.

We now compute the coefficient $\cal C$ for the process $\Theta^{+} \to W^{+}g$ by 
evaluating the 1-loop diagrams 
 \bear\
 \!\! \hspace*{-0.85cm} 
{\cal M} = \int \!\!\frac{d^{4} k}{(2\pi)^{4}} && \hspace*{-0.7cm}  
\frac{ 48 i \, \mu_{\Theta} g_{s}  g\,}{(k^2\!-\! M_{\varphi}^2) [\left(p_1\! + \! p_2 \! + \! k \right)^2 \!-\! M_{\varphi}^2 ] }\! 
\nonumber \\ [3mm]
&&  \hspace*{-1.5cm} 
 \times\left(  \frac{1}{4}\, \epsilon_{1}\!\cdot \epsilon_{2} 
 -   \frac{ \epsilon_{1}\!\cdot\!\! \left(p_{2} + k \right) (\epsilon_{2}\!\cdot \! k)}{ \left(p_{2}\! +\! k \right)^2 \! -\! M_{\varphi}^2 } 
\right)  ~, 
\label{amplitude}
 \eear
where $k$ is the loop 4-momentum and $M_{\varphi}$ is the mass of the 
scalars running in the loop.
The logarithmic divergences from the three diagrams  cancel, and
  $\cal C$ in Eq.\,(\ref{amplitude2}) can be written as a  Feynman parameter integral
\be 
{\cal C} =  \int_{0}^{1} \!\!dx  \int_{0}^{1-x} \!\!\!\!\! dy\,\frac{ -3(1-R^{2}) \,x y}{1 - x y - R^{2}\, x \,(1 - x - y)} ~.\quad
\ee

\noindent with $R \equiv M_{W}/M_{\varphi}$. After integration over $y$ we obtain
\be
{\cal C} = \!\frac{-3}{2(1 \!-\! R^{2})}\!\left[ \frac{\pi^2}{9} \!-\! 1 
+ R^{2} \! \left( \!\frac{\pi}{\sqrt{3}} \!-\! 1 \!\right) \!+\! 2 J(R) \right] ,\>\>\>
\label{coeff}
\ee
where we have defined
\be
\!\!J(R) =  \!\int^1_0 \!\! d x \left(\frac{1}{x} \!-\! R^{2} \,x\! \right) \ln \left[1 \!-\! R^{2} \,x (1\!-\! x) \right] ~.
\ee
For $R\ll 1$ the function $J$ has the form
\be
J(R) = -\frac{R^{2}}{2}\left(1-\frac{R^{2}}{12} - \frac{R^4}{180}\right) + \mathcal O(R^{8}) ~~.
\ee

After squaring the amplitude (\ref{amplitude2}) and summing over final state polarizations,  we find the 
following decay width:
\be
\Gamma (\Theta^{+}\!\to W^{+} g)= \frac{\alpha_{s} \alpha \, \mu_{\Theta}^{2} }{\pi^{3}  \sin^{2}\!\theta_{W} M_\Theta} f(R) ~~,
\label{generic-width}
\ee
where we have assumed $M_{\varphi} = M_{\Theta}$, as is the case for the octo-triplet (see 
Figure\,\ref{fig:triangle}), and 
\be
f(R) =  \frac{1}{2}\, \mathcal{C}^{2} (1-R^{2})~~,
\label{function}
\ee
with $\cal C$ depending on $R$ as shown in Eq. (\ref{coeff}).
This function, which  appears in all 1-loop decays discussed in this paper,
is accidentally suppressed by cancellations between terms involving various powers of $\pi$.
To see this, consider the  expansion around $R\equiv M_W/M_\Theta \to 0$:
\be
f(R)= f(0) + f_1 R^{2} + f_2 R^4 +  \mathcal O (R^{6}) ~~,
\label{function-expansion}
\ee
Each of the above coefficients happens to be much smaller than order one: 
\bear
f(0) &\!\!=\!\!& \frac{9}{8}\!\left(\frac{\pi^{2}}{9}-1\right)^{\!2} \simeq  \,1.05 \times 10^{-2}   ~~,
\nonumber \\ [2mm]
f_1 &\!\!=\!\!& f(0) + \frac{9}{4}\left(\frac{\pi^{2}}{9}-1\right) \left(\frac{\pi}{\sqrt{3}} - 2 \right)
\nonumber \\ [2mm]
 &\!\!\simeq\!\!&  -3.00 \times 10^{-2}  ~~,
\nonumber \\ [3mm]
f_2 &\!\!=\!\!& \frac{9}{8}\!\left(\frac{\pi^{2}}{9} + \frac{\pi}{\sqrt{3}} -3 \! \right)^{\!\! 2} + \frac{3}{16}\left(\!\frac{\pi^{2}}{9}-1\!\right)
\nonumber \\ [2mm]
&\!\!\simeq\!\!&  2.71 \times 10^{-2}  ~~.\label{flimit}
\eear 
The above value of $f(0)$ agrees with that extracted 
from the width of an octo-doublet \cite{Gresham:2007ri} or octo-singlet \cite{Dobrescu:2007yp} decaying into $gg$.

Eqs.~(\ref{generic-width}), (\ref{function-expansion}) and (\ref{flimit}) show that the 2-body decays of the octo-triplet into gauge bosons 
are suppressed by two orders of magnitude compared to estimates based on dimensional analysis.

 \vfil 

\begin{thebibliography}{99} \frenchspacing

\bibitem{Chivukula:1991zk}
  R.~S.~Chivukula, M.~Golden and E.~H.~Simmons,
  ``Multi - jet physics at hadron colliders,''
  Nucl.\ Phys.\  B {\bf 363}, 83 (1991).

\bibitem{Dobrescu:2007yp}
  B.~A.~Dobrescu, K.~Kong and R.~Mahbubani,
  ``Massive color-octet bosons and pairs of resonances at hadron colliders,''
  Phys.\ Lett.\  B {\bf 670}, 119 (2008)
  [arXiv:0709.2378 [hep-ph]].

\bibitem{Bai:2010dj}
  Y.~Bai, B.~A.~Dobrescu,
  ``Heavy octets and Tevatron signals with three or four $b$ jets,''
  [arXiv:1012.5814 [hep-ph]].

\bibitem{HiggsPlusb}
CDF Collaboration, ``Search for Higgs bosons produced in association with $b$ quarks'',
Note 10105, June 2010, \url{http://www-cdf.fnal.gov/physics/new/hdg//}
\url{Results_files/results/3b_susyhiggs_jun10}

\bibitem{Boughezal:2010ry}
  R.~Boughezal, F.~Petriello,
  ``Color-octet scalar effects on Higgs boson production in gluon fusion,''
  Phys.\ Rev.\  {\bf D81}, 114033 (2010).
  [arXiv:1003.2046 [hep-ph]]. \\
  R.~Boughezal,
  ``Constraints on heavy colored scalars from Tevatron's Higgs exclusion limit,''
  [arXiv:1101.3769 [hep-ph]].

\bibitem{Hill:2002ap} For a review, see 
  C.~T.~Hill and E.~H.~Simmons,
  ``Strong dynamics and electroweak symmetry breaking,''
  Phys.\ Rept.\  {\bf 381}, 235 (2003)
  [Erratum-ibid.\  {\bf 390}, 553 (2004)]
  [arXiv:hep-ph/0203079].

\bibitem{Kilic:2008pm}
  C.~Kilic, T.~Okui and R.~Sundrum,
  ``Colored Resonances at the Tevatron: Phenomenology and Discovery Potential
  in Multijets,''
  JHEP {\bf 0807}, 038 (2008)
  [arXiv:0802.2568 [hep-ph]].
``Vectorlike Confinement at the LHC,''
  JHEP {\bf 1002}, 018 (2010)
  [arXiv:0906.0577 [hep-ph]].

\bibitem{Bai:2010mn}
  Y.~Bai and A.~Martin,
  ``Topological Pions,''
  Phys.\ Lett.\  B {\bf 693}, 292 (2010)
  [arXiv:1003.3006]. \\ 
  Y.~Bai and R.~J.~Hill,
  ``Weakly interacting stable hidden sector pions,''
  Phys.\ Rev.\  D {\bf 82}, 111701 (2010)
  [arXiv:1005.0008 [hep-ph]].

\bibitem{Burdman:2006gy}
  G.~Burdman, B.~A.~Dobrescu and E.~Ponton,
  ``Resonances from two universal extra dimensions,''
  Phys.\ Rev.\  D {\bf 74}, 075008 (2006)
  [arXiv:hep-ph/0601186].

\bibitem{Martynov:2009en}
  M.~V.~Martynov and A.~D.~Smirnov,
  ``Chiral color symmetry and possible $G'$-boson effects at the Tevatron and LHC,''
  Mod.\ Phys.\ Lett.\  A {\bf 24}, 1897 (2009)
  [arXiv:0906.4525 [hep-ph]].

\bibitem{Manohar:2006ga}
  A.~V.~Manohar and M.~B.~Wise,
  ``Flavor changing neutral currents, an extended scalar sector, and the  Higgs
  production rate at the LHC,''
  Phys.\ Rev.\  D {\bf 74}, 035009 (2006)
  [arXiv:hep-ph/0606172].

\bibitem{Arnold:2009ay}
  J.~M.~Arnold, M.~Pospelov, M.~Trott and M.~B.~Wise,
  ``Scalar representations and minimal flavor violation,''
  JHEP {\bf 1009}, 073 (2010).
  [arXiv:0911.2225 [hep-ph]].

\bibitem{Gresham:2007ri}
  M.~I.~Gresham, M.~B.~Wise,
  ``Color octet scalar production at the LHC,''
  Phys.\ Rev.\  {\bf D76}, 075003 (2007).
  [arXiv:0706.0909 [hep-ph]].

\bibitem{Gerbush:2007fe}
  M.~Gerbush {\it et al.},  
  ``Color-octet scalars at the LHC,''
  Phys.\ Rev.\  {\bf D77}, 095003 (2008).
  [arXiv:0710.3133]. 

\bibitem{CDF7}
CDF Collaboration, 
\url{http://www-cdf.fnal.gov/} \url{physics/ewk/2011/wjj/7_3.html}, May 2011.

\bibitem{Aaltonen:2011mk}
  T.~Aaltonen {\it et al.}  [CDF Collaboration],
  ``Invariant Mass Distribution of Jet Pairs Produced in Association with a $W$
  boson in $p \bar{p}$ Collisions at $\sqrt{s}= 1.96$ TeV,''
  arXiv:1104.0699. 

\bibitem{Abazov:2011af}
  V.~M.~Abazov {\it et al.}  [D0 Collaboration],
  ``Study of the dijet invariant mass distribution in $p\bar{p}\to
  W(\to\ell\nu)+ {jj}$ final states at $\sqrt{s} =1.96$ TeV,''
  arXiv:1106.1921 [hep-ex].

\bibitem{Yu:2011cw}
  F.~Yu,
  ``A $Z^\prime$  model for the CDF dijet anomaly,''
  arXiv:1104.0243. \\
  M.~R.~Buckley, D.~Hooper, J.~Kopp and E.~Neil,
  ``Light Z' bosons at the Tevatron,''
  arXiv:1103.6035. \\
  K.~Cheung, J.~Song,
  ``Tevatron Wjj Anomaly and the baryonic $Z'$ solution,''
  arXiv:1104.1375. \\
  X. P.~Wang {\it et al,} 
  ``New color-octet vector boson revisit,''
  arXiv:1104.1917;
``O(100 GeV) deci-weak $W^\prime/Z^\prime$ at Tevatron and LHC,''
  arXiv:1104.1161. \\
  L.~A.~Anchordoqui  {\it et al,} 
  ``Stringy origin of Tevatron Wjj anomaly,''
  arXiv:1104.2302.


\bibitem{Nelson:2011us}
  A.~E.~Nelson, T.~Okui, T.~S.~Roy,
  ``A unified, flavor symmetric explanation for the $t\bar{t}$ asymmetry and Wjj
  excess at CDF,''  arXiv:1104.2030. \\
  C.~Kilic and S.~Thomas,
  ``Signatures of resonant super-partner production with charged-current decays,''
  arXiv:1104.1002.\\
  E.~J.~Eichten, K.~Lane and A.~Martin,
  ``Technicolor at the Tevatron,''
  arXiv:1104.0976.



\bibitem{Sato:2011ui}
  R.~Sato, S.~Shirai, K.~Yonekura,
  ``A Possible Interpretation of CDF Dijet Mass Anomaly and its Realization in Supersymmetry,''
  arXiv:1104.2014.


\bibitem{Dodelson:1991iv}
  S.~Dodelson, B.~R.~Greene, L.~M.~Widrow,
  ``Baryogenesis, dark matter and the width of the Z,''
  Nucl.\ Phys.\  {\bf B372}, 467-493 (1992). \\ 
  M.~Cirelli, N.~Fornengo, A.~Strumia,
  ``Minimal dark matter,''
  Nucl.\ Phys.\  {\bf B753}, 178-194 (2006).
  [hep-ph/0512090].


\bibitem{Alwall:2011uj}
  J.~Alwall, M.~Herquet, F.~Maltoni, O.~Mattelaer, T.~Stelzer,
  ``MadGraph 5 : Going Beyond,''
  [arXiv:1106.0522 [hep-ph]].

\bibitem{Christensen:2008py}
  N.~D.~Christensen, C.~Duhr,
  ``FeynRules - Feynman rules made easy,''
  Comput.\ Phys.\ Commun.\  {\bf 180}, 1614-1641 (2009).
  [arXiv:0806.4194]; 
FeynRules, version 1.6beta, \url{http://feynrules.irmp.ucl.ac.be/}; \\
Model files for heavy octets are available at \url{http://theory.fnal.gov/people/dobrescu/octet/}.

\bibitem{Pumplin:2002vw}
  J.~Pumplin {\it et al.}, 
  ``New generation of parton distributions with uncertainties from global QCD
  analysis,''
  JHEP {\bf 0207}, 012 (2002)
  [arXiv:hep-ph/0201195].

\bibitem{Kats:2009bv}
  Y.~Kats, M.~D.~Schwartz,
  ``Annihilation decays of bound states at the LHC,''
  JHEP {\bf 1004}, 016 (2010).
  [arXiv:0912.0526 [hep-ph]].

\bibitem{Kim:2008bx}
  C.~Kim, T.~Mehen,
  ``Color Octet Scalar Bound States at the LHC,''
  Phys.\ Rev.\  {\bf D79}, 035011 (2009).
  [arXiv:0812.0307 [hep-ph]]. \\
  A.~Idilbi, C.~Kim, T.~Mehen,
  ``Pair Production of Color-Octet Scalars at the LHC,''
  Phys.\ Rev.\  {\bf D82}, 075017 (2010).
  [arXiv:1007.0865].
 
    
\bibitem{Gamiz:2009ku}
  E.~Gamiz {\it et al.} 
                  [HPQCD Collaboration],
  ``Neutral $B$ meson mixing in unquenched lattice QCD,''
  Phys.\ Rev.\  D {\bf 80}, 014503 (2009)
  [arXiv:0902.1815 [hep-lat]].
  
\bibitem{Buras:2001ra}
  A.~J.~Buras, S.~Jager and J.~Urban,
  ``Master formulae for $\Delta F = 2$ NLO-QCD factors in the standard model  and
  beyond,''
  Nucl.\ Phys.\  B {\bf 605}, 600 (2001)
  [arXiv:hep-ph/0102316].
  
\bibitem{Becirevic:2001xt}
  D.~Becirevic  {\it et al.},  
  ``$B$-parameters of the complete set of matrix elements of $\Delta B = 2$
  operators from the lattice,''
  JHEP {\bf 0204}, 025 (2002)
  [arXiv:hep-lat/0110091].
  
\bibitem{Lenz:2006hd}
  A.~Lenz and U.~Nierste,
  ``Theoretical update of $B_s - \bar{B}_s$ mixing,''
  JHEP {\bf 0706}, 072 (2007)
  [arXiv:hep-ph/0612167].


\bibitem{Dobrescu:2010rh}
  B.~A.~Dobrescu, P.~J.~Fox, A.~Martin,
  ``CP violation in $B_s$ mixing from heavy Higgs exchange,''
  Phys.\ Rev.\ Lett.\  {\bf 105}, 041801 (2010).
  [arXiv:1005.4238 [hep-ph]].
 
\bibitem{Abazov:2010hv}
  V.~M.~Abazov {\it et al.}  [D0 Collaboration],
  ``Evidence for an anomalous like-sign dimuon charge asymmetry,''
  Phys.\ Rev.\  D {\bf 82}, 032001 (2010)
  [arXiv:1005.2757 [hep-ex]].

\bibitem{6jet}
CDF Collaboration, ``Search for a new hadronic resonance using jet ensembles with CDF'',
note 10256, February 2011.

 \bibitem{pythia}
  T.~Sjostrand, S.~Mrenna and P.~Skands,
  ``{\sc Pythia} 6.4 physics and manual,''
  JHEP {\bf 0605}, 026 (2006).
  [arXiv:hep-ph/0603175].
  
  \bibitem{pgs}
J.~S.~Conway, ``Pretty Good Simulation of high-energy collisions'', 090401 release,
\url{http://physics.ucdavis.edu/~conway/} \hspace{-0.2cm} research/  \\ software/pgs/pgs4-general.htm

\bibitem{chameleon}
J.~Thaler {\it et al},  Chameleon version 1.02, July 2006,
\url{http://v1.jthaler.net/olympicswiki/}
\url{doku.php?id=lhc_olympics:analysis_tools}.

\bibitem{Abazov:2008gc}
  V.~M.~Abazov {\it et al.}  [D0 Collaboration],
  ``Measurement of the $t \bar{t}$ production cross section in $p \bar{p}$
  collisions at $\sqrt{s}$ = 1.96 TeV,''
  Phys.\ Rev.\ Lett.\  {\bf 100}, 192004 (2008)
  [arXiv:0803.2779]. \\
  T.~Aaltonen {\it et al.}  [CDF Collaboration],
  ``First measurement of the ratio $\sigma_{t\bar{t}} / \sigma_{Z/\gamma^* \to \ell\ell }$ and
  precise extraction of the $t\bar{t}$ cross section,''
  Phys.\ Rev.\ Lett.\  {\bf 105}, 012001 (2010)
  [arXiv:1004.3224].
   

\bibitem{Hill:1991at}
  C.~T.~Hill,
  ``Topcolor: Top quark condensation in a gauge extension of the standard model,''
  Phys.\ Lett.\  {\bf B266}, 419-424 (1991).

\bibitem{Hall:1985wz}
  L.~J.~Hall and A.~E.~Nelson,
  ``Heavy gluons and monojets,''
  Phys.\ Lett.\  B {\bf 153}, 430 (1985).

\bibitem{CDF7kin}
CDF Collaboration, 
\url{http://www-cdf.fnal.gov/} \url{physics/ewk/2011/wjj/kinematics.html}, May 2011.


\end{thebibliography}
\end{document}